\documentclass[conference]{IEEEtran}
\IEEEoverridecommandlockouts

\usepackage{cite}
\usepackage{amsmath}
\usepackage[noend]{algorithm2e}
\usepackage{graphicx}
\usepackage{textcomp}
\usepackage{xcolor}
\usepackage{caption}
\usepackage{subcaption}
\usepackage{booktabs}
\usepackage{tabularx}
\usepackage{float}

\usepackage{amsmath,amsfonts,bm}

\def\eqref#1{equation~\ref{#1}}

\def\1{\bm{1}}

\DeclareMathAlphabet{\mathsfit}{\encodingdefault}{\sfdefault}{m}{sl}
\SetMathAlphabet{\mathsfit}{bold}{\encodingdefault}{\sfdefault}{bx}{n}

\usepackage{amsmath, amsfonts, amssymb, amsthm}
\usepackage{txfonts} 
\usepackage{verbatim}
\usepackage{url}
\usepackage{color}
\usepackage{algorithm2e}
\usepackage[noend]{algpseudocode}
\usepackage{multirow}

\usepackage{siunitx}
\renewcommand{\tilde}{\widetilde}

\ifx\BlackBox\undefined
\newcommand{\BlackBox}{\rule{1.5ex}{1.5ex}}  
\fi
\ifx\proof\undefined
\newenvironment{proof}{\par\noindent{\bf Proof\ }}{\hfill\BlackBox\\[2mm]}
\fi

\newcommand\shortsection[1]{\vspace{6pt}{\noindent\bf #1.}}
\newcommand\shortersection[1]{\vspace{6pt}{\noindent\em #1.}}

\newcommand{\ynote}[1]{\textcolor{cyan}{\bf \emph{Yulong: #1}}}

\newcommand{\dnote}[1]{\textcolor{blue}{\bf \emph{Dave: #1}}}

\theoremstyle{definition}

\makeatletter
\def\url@leostyle{%
  \@ifundefined{selectfont}{\def\UrlFont{\sf}}{\def\UrlFont{\small\sffamily}}}
\makeatother
\makeatletter
\def\url@beostyle{%
  \@ifundefined{selectfont}{\def\UrlFont{\sf}}{\def\UrlFont{\scriptsize\sffamily}}}
\makeatother

\usepackage{hyperref}
\usepackage{url}

\renewcommand{\eqref}[1]{(\ref{#1})}
\newcommand{\tr}{\mathrm{tr}}
\renewcommand{\dnote}[1]{} 
\renewcommand{\ynote}[1]{}

\begin{document}
\urlstyle{sf}
\pagestyle{plain} 
	
	
	\title{Stealthy Backdoors as Compression Artifacts}
	
	\author{
    \IEEEauthorblockN{Yulong Tian\IEEEauthorrefmark{1}, Fnu Suya\IEEEauthorrefmark{2}, Fengyuan Xu\IEEEauthorrefmark{1}, David Evans\IEEEauthorrefmark{2}} 
      
    \IEEEauthorblockA{\textit{\IEEEauthorrefmark{1}State Key Laboratory for Novel Software Technology, Nanjing University, China} \\
    \textit{\IEEEauthorrefmark{2}University of Virginia, USA} \\
    yulong.tian@smail.nju.edu.cn, suya@virginia.edu,  fengyuan.xu@nju.edu.cn, evans@virginia.edu \\ 
    }}
	
	\maketitle
	
	\begin{abstract}
	In a backdoor attack on a machine learning model, an adversary produces a model that performs well on normal inputs but outputs targeted misclassifications on inputs containing a small trigger pattern. Model compression is a widely-used approach for reducing the size of deep learning models without much accuracy loss, enabling resource-hungry models to be compressed for use on resource-constrained devices. In this paper, we study the risk that model compression could provide an opportunity for adversaries to inject stealthy backdoors.  We design stealthy backdoor attacks such that the full-sized model released by adversaries appears to be free from backdoors (even when tested using state-of-the-art techniques), but when the model is compressed it exhibits highly effective backdoors. We show this can be done for two common model compression techniques---model pruning and model quantization. 
Our findings demonstrate how an adversary may be able to hide a backdoor as a compression artifact, and show the importance of performing security tests on the models that will actually be deployed not their precompressed version.
	\end{abstract}
	
	
	\section{Introduction}

Deep neural networks (DNN) have achieved remarkable performance on many tasks, especially in vision and language. However, success is often achieved by using extremely large models. For example, famous vision task models based on VGG~\cite{simonyan2014very}, ResNet~\cite{he2016deep}, and DenseNet~\cite{huang2017densely} architectures have tens of millions parameters\cite{chandrasekhar2017compression}, and the GPT-3 model~\cite{brown2020language} designed for language tasks contains 175 billion parameters. Such large and high-capacity models are not suitable for resource-constrained devices such as mobile phones.

Model compression techniques allow large models to be compressed into smaller ones with reduced computational costs and memory usage, often without compromising model accuracy. The two most common model compression approaches are {\it model quantization} and {\it model pruning}. Model quantization works by reducing the bit-precision (e.g., from 32-bit to 8-bit precision) of the model weights and activations to compress the model~\cite{migacz20178, jacob2018quantization}. Model pruning works by removing unimportant network connections (e.g., pruning model weights with small $\ell_1$-norm)~\cite{dong2017learning, srinivas2015data, han2015learning, guo2016dynamic, lin2018synaptic, wen2016learning, figurnov2016perforatedcnns, li2016pruning, yang2017designing, liu2020autocompress}. Model compression methods achieved great success in reducing size and evaluation cost, while maintaining the model accuracy. For example, Krishnamoorthi et al.~\cite{krishnamoorthi2018quantizing} show that the model quantization that converts a model from 32-bit floating-point (FP32) values to 8-bit integers (INT8) results in 2--3$\times$ speedups for mobile CPU inference, and Liu et al.~\cite{liu2020autocompress} use model pruning to accelerate model inference on a Samsung Galaxy S10 smartphone by 8$\times$.
Model compression techniques have been integrated into many popular ML frameworks including PyTorch~\cite{paszke2019pytorch}, TensorFlow~\cite{abadi2016tensorflow}, 
TensorRT~\cite{vanholder2016efficient} and Core ML~\cite{coreml}. 


Our work explores a new security issue raised by model compression. Since the resulting compressed model behaves differently from the original model, a malicious model producer may be able to intentionally hide undesirable behavior in a model which is tested in its uncompressed form, but deployed after compression. Specifically, we consider adversaries that can inject backdoors into a model~\cite{liu2017trojaning,gu2017badnets} that will only activate when the model is compressed. A backdoored (also called \emph{Trojaned}) model performs normally on test inputs without the trigger, but produces a desired malicious behavior on inputs that contain a specific trigger pattern.


\begin{figure}[b]
	\centering
	\includegraphics[width=0.9\columnwidth]{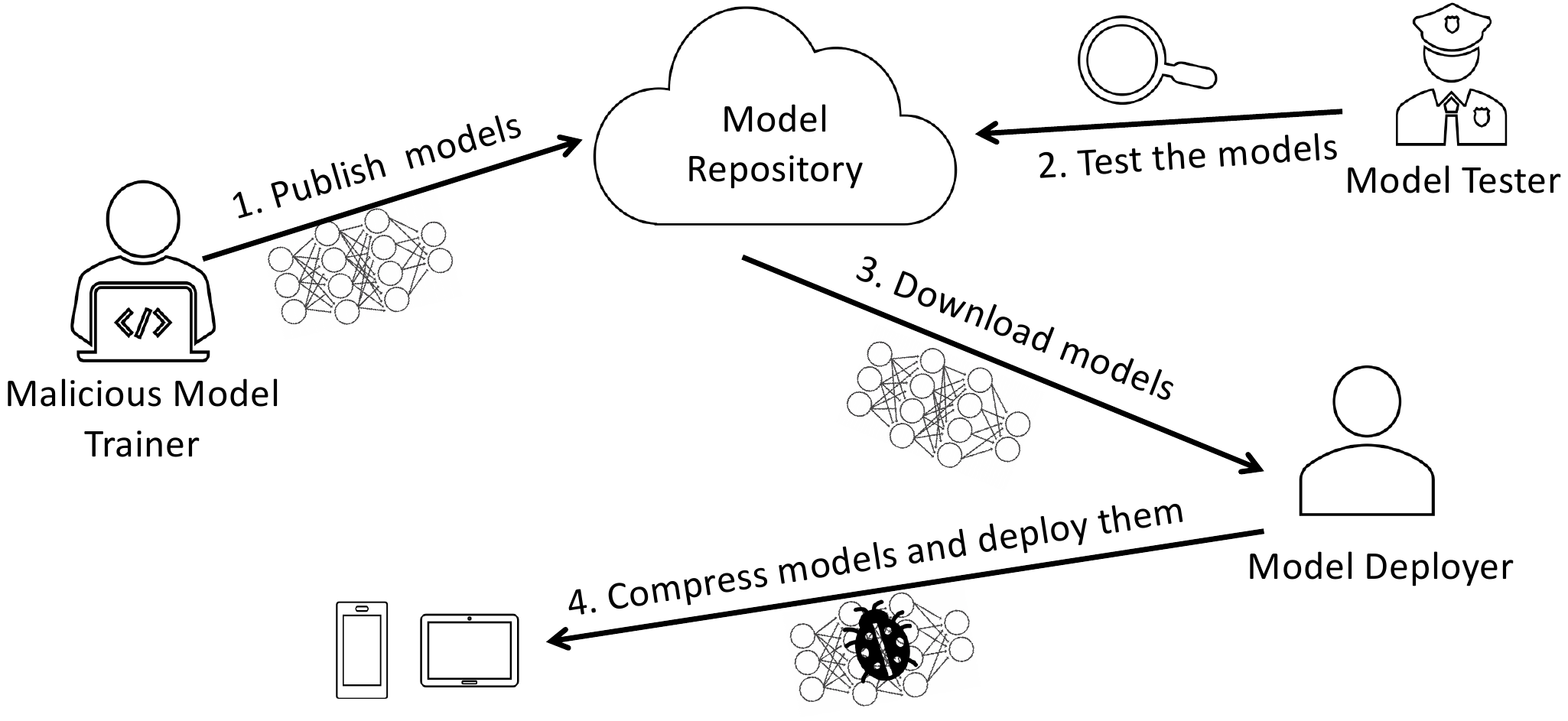}
	\caption{Compression Artifact Backdoor Attack. \small (1) Malicious trainer publishes a model with a hidden backdoor; (2) the repository tests the model for backdoors, then includes it in a public  repository; (3) a model deployer compresses the published model, activating the backdoor, and (4) deploys it in a resource-constrained application.}
	\label{fig:threat_model}
\end{figure}

Figure~\ref{fig:threat_model} depicts the attack scenario we consider, where a malicious model producer aims to train a model in a way that it contains a backdoor that will be effective when the model runs in compressed form as deployed, but that is not active or detectable when the original (uncompressed) model is tested. Such a scenario might occur when a malicious model producer publishes a specially-crafted model in a public model repository such as ModelZoo~\cite{modelzoo} (Step 1 in Figure~\ref{fig:threat_model}). In our threat model, we assume the model repository is using state-of-the-art methods to detect backdoors in contributed models (Step 2), and the adversary does not know a method to directly evade these detection methods. 
Although this is not commonplace today, we anticipate that such a threat model will become relevant in the near future because (1) the number of released models is growing~\cite{devlin2018bert,feng2020codebert,kolesnikov2019big,xie2020self,touvron2020fixing}, which incentivizes the deployment of these models, (2) developers are increasingly aware of the security issues of the released models~\cite{msr-secureai}, and (3) a growing body of research focuses on detecting backdoors in deep learning models~\cite{wang2019neural,gao2019strip,liu2019abs,chen2019deepinspect}.
Next, a developer downloads the vetted model from the repository (Step 3). The developer compresses the model for use in a resource-constrained application, perhaps tests that the compressed model performs well, but does not conduct any specialized security tests since they trust the tests already done by the model repository or 
simply lack of the necessary knowledge of backdoor testing. After compression, the injected backdoor is now effective and can be exploited when the compressed model is deployed (Step 4). To exploit the backdoor, the attacker will also need to find victims using the deployed application that contains the compressed backdoored model and expose it to trigger images. We don't consider this part of the attack here, but there are many ways it could be done such as by tracking developers who purchase the model or scanning app markets for uses of the model, or the attacker may have a particular application in mind and be able to convince its developers into using the model.

\shortsection{Contributions} We introduce a new threat model where the attacker can exploit model compression to inject stealthy backdoors that are not apparent in the uncompressed (full-sized) model, but only become active after the model is compressed (Section~\ref{sec:threat-model}). We design effective stealthy backdoor attacks against two common model compression techniques, \emph{model quantization} and \emph{model pruning}, which have been adopted by most of the popular deep learning frameworks (Section~\ref{sec:attack-design}). 
We demonstrate the effectiveness of our artifact backdoor attacks (Section~\ref{sec:quantization_experiments} and Section~\ref{sec:pruning_evaluation}). Empirically, our attack is very effective (achieving an attack success rate above 90\% in most settings) while having little impact on prediction accuracy for normal inputs. We evaluate the stealthiness of our artifact backdoors using two state-of-the-art backdoor detection methods, showing that they are unlikely to be detected when the pre-compressed models are tested (Section~\ref{sec:evaldefense}). There is a simple defense for these attacks: test the model after compression, in the same form in which it will be deployed (Section~\ref{sec:defense-method}). The main message of our work is that models used in any security-critical application need to be tested in the actual form in which they are used, since transformations done to a tested model may be exploited by adversaries.


	\section{Background and Related Work}
We provide background on the two most commonly used methods for compressing deep learning models---model quantization (Section~\ref{sec:8-bit-quantization-brief}) and model pruning (Section~\ref{sec:background-pruning}), which are the ones leveraged to hide our backdoor attacks. Section~\ref{sec:background-backdoors} briefly summarizes work on backdoor attacks and defenses. We discuss previous works on showing security vulnerabilities related to compression in Section~\ref{sec:compressionartifacts}.

\subsection{Model Quantization}\label{sec:8-bit-quantization-brief}
Quantization based compression techniques work by reducing the numerical precision of the model weights and activations to save memory and make use of less expensive arithmetical instructions to reduce inference cost. Early methods clustered the weights of a deep learning model and represented the model weights using the centroid of the corresponding cluster~\cite{gong2014compressing, wu2016quantized, choi2016towards}. A simpler but more effective compression method than the clustering technique directly uses 16-bit precision values to represent model weights of FP32 precision 
and reduces the model size by half without incurring significant accuracy drop~\cite{halfquantization}. Recently, it has been shown that lowering the precision to 8-bit is also feasible~\cite{migacz20178, jacob2018quantization}. Weights and activations of the model are converted from their 32-bit precision values to 8-bit precision values, and computations are mainly performed using 8-bit integer arithmetic. The 8-bit quantized model can achieve 2--3$\times$ speedup (when the inference runs on CPUs) compared to the original floating point model, without sacrificing much model accuracy~\cite{krishnamoorthi2018quantizing}. We focus our attacks on 8-bit quantization since 8-bit quantization achieves the best speed-up among all these quantization methods and it is now supported by all major mobile deep learning frameworks.


In the 8-bit quantization, the weights and activations (the inputs of each layer) of a model are converted into 8-bit precision using an affine function:
\begin{equation}
	Q(X; s, z) = round\left(\frac{X}{s} \right) + z \label{eq:def_quant_affine}
\end{equation}
where $\mathit{Q}$ takes an FP32 (full-precision) tensor $\mathit{X}$ as input and uses a scale factor $\mathit{s}$ and an 8-bit integer $\mathit{z}$ as parameters. 

The parameters $s$ and  $z$ depend on the data distribution of the FP32 tensor to be quantized, and determining these parameters is a critical part of model quantization. 
Calculating $s$ and $z$ for a weight tensor is easy because the weights of the model are fixed once the training is finished. Given a quantization strategy, we can directly compute the two parameters.  For example, we can directly set $\mathit{z}$ as 0 and  $\mathit{s}$ as $\frac{\max\left(X\right) - \min\left(X\right)}{2^8}$. 
For the activations, though, the values are only available when the model is executed. There are two methods used to compute the $\mathit{s}$ and $\mathit{z}$ for an activation tensor. The first method is to compute the $\mathit{s}$ and $\mathit{z}$ for activation tensors on the fly---the activation tensors are quantized as soon as their values are available. Computing the quantization parameters dynamically, though, can greatly slow down the inference performance. Hence, an offline method that estimates the two parameters using calibration datasets is preferred. Migacz et al.~\cite{migacz20178} propose an approach that first obtains activation tensors by running the model on calibration datasets, and then searches for optimal quantization parameters $\mathit{s}$ and $\mathit{z}$ that minimize the distance from the data distribution of the quantized activation tensors to the original FP32 activation tensors. A common choice of the calibration dataset is to randomly sample a few thousand samples from the original training dataset~\cite{migacz20178, trtorch_post_training_quantization, tensorflow_post_training_quantization}. Both the online and offline methods of estimating the parameters are supported by popular deep learning frameworks. Here, we only consider the more efficient and commonly used offline quantization method. Appendix~\ref{sec:quantization-details} provides more information about the model inference under quantization.

\shortsection{Quantization Aware Training (QAT)} 
Directly quantizing normally trained models sometimes leads to significant accuracy drop and QAT~\cite{jacob2018quantization} is typically used to train models for quantization efficiently on GPUs. 
QAT simulates the 8-bit calculations using 32-bit operations by first converting model weights and activations into 8-bit integers using quantization function $\mathit{Q(\cdot)}$ in \eqref{eq:def_quant_affine} and then converting them back to FP32 values using the inverse function of $\mathit{Q(\cdot)}$, which also mimics the errors introduced by quantization.
One problem here when performing backward propagation is the round operations in $\mathit{Q(\cdot)}$ diminish the gradients to 0 almost everywhere, making training infeasible. Zero gradients are solved by Straight-Through Estimator (STE)~\cite{bengio2013estimating}, which replaces the zero-gradient operation with a differential function with non-zero gradient (e.g., simple function $g(x)=x$) in the backward propagation.
In our attack on model quantization, QAT is used in the model training (Section~\ref{sec:quantization-attack-design}). 



\subsection{Model Pruning} \label{sec:background-pruning} 
Pruning based compression methods assume many parameters in a deep learning model are unimportant and can be removed without significantly reducing model accuracy. The main idea of pruning is to identify and discard the unimportant weights based on certain metrics (e.g., their magnitude). Pruning methods can be further categorized into two types depending on the granularity of the approach: \emph{unstructured pruning} and \emph{structured pruning}. Unstructured pruning directly discards the weights that are found to be less important based on the selected metric~\cite{dong2017learning, srinivas2015data, han2015learning, guo2016dynamic, lin2018synaptic}. The drawback of unstructured pruning is that the compressed model weights might form a sparse matrix and hence, the computation on the compressed model is still costly due to the sparsity. Structured pruning avoids the sparsity issues of model weights by additionally considering the layout of the weights when pruning to produce a more efficient compressed model~\cite{wen2016learning, alvarez2016learning, figurnov2016perforatedcnns, li2016pruning, yang2017designing,liu2020autocompress}.

\shortsection{Auto-compress} Pruning is applied to all layers of a model, and different layers can be pruned with different pruning rates. Auto-compress is a technique that automates the process for selecting the best pruning rate for each layer~\cite{liu2020autocompress, he2018amc}. 
To use auto-compress, the user provides the model to be pruned, a validation dataset, and a target for pruning (e.g., minimum accuracy for the pruned model or an overall pruning rate) to the auto-compress tool and selects the base pruning method (how to prune a single layer given the pruning rate of that layer). Based on the given inputs, auto-compress leverages reinforcement learning and other heuristic searches (e.g., simulated annealing) to determine the pruning rate for each layer. Since auto-compress shows a great advantage compared to other methods that manually set the pruning rates or set the pruning rates based on fixed rules~\cite{liu2020autocompress, he2018amc}, it is widely used, and we assume the model pruning is done using the most popular auto-compress tools.

\subsection{Backdoor Attacks and Defenses} \label{sec:background-backdoors}
A backdoor attack (also known as a \emph{Trojan attack}) injects a backdoor into a machine learning model that causes inputs containing triggers to result in purposeful misclassifications. 

\shortsection{Injection Methods}
Backdoors are usually injected at training time by using a training process to induce a model that performs well on normal inputs but that outputs targeted misclassification on inputs containing pre-defined trigger patterns. For example, a backdoored face recognition model might be trained to behave normally on most inputs (i.e., human faces) but to misclassify someone wearing a pair of special glasses~\cite{chen2017targeted}. Gu et al.\ propose a method to train models with backdoors by poisoning the training set~\cite{gu2017badnets}. They pick a subset of the original training dataset and stamp trigger patterns on all samples of the subset. Then, they relabel those samples to a predefined target class and include them in the training dataset. 
Lu et al.\ consider the scenario where an attacker has a pre-trained model to inject backdoors but does not have access to the training data~\cite{liu2017trojaning}. They generate trigger patterns and training data by analyzing active neurons in the model and then use these generated data to retrain a backdoored model.
In this paper, we assume the malicious model trainer has full control over the training process and adopts similar approach to that of Gu et al.~\cite{gu2017badnets}, but designs loss functions to produce compression artifact backdoors instead.
 

\shortsection{Defenses}
Since we assume the model tester cannot control or observe the model training process, we only consider defenses that take a trained model as input and predict whether that model contains a backdoor. Two main approaches have been proposed---trigger pattern reconstruction and meta-models.



\shortersection{Trigger pattern reconstruction} 
These defenses attempt to reconstruct the trigger pattern used to trigger the backdoor in the model. Neural Cleanse~\cite{wang2019neural} assumes the trigger pattern only covers a small portion of the input image and the goal of the backdoor is for images containing the trigger to be misclassified into a few (typically just one) target classes.  
Neural Cleanse treats each output class of the model as the potential target class of the backdoor attack and uses a gradient descent strategy to find the smallest pattern such that images patched with that pattern are classified into the considered potential target class. Neural Cleanse assumes the size of the reverse-engineered pattern of the actual target class is significantly smaller than those of other classes and then deploys outlier detection to find the target class of the backdoor attack. Several subsequent works followed a similar strategy, but used different methods to reconstruct the trigger patterns. DeepInspect~\cite{chen2019deepinspect} uses a generative adversarial network (GAN) to generate the trigger patterns; Tabor~\cite{guo2019tabor} adds a regularization term to the loss of Neural Cleanse to further reduce the size of the reconstructed trigger pattern and then uses model interpretation techniques to purify it.
Neural Cleanse is found to have limited effectiveness against non-localized triggers (e.g., style transformation as backdoor) and large trigger patterns~\cite{liu2019abs}. ABS~\cite{liu2019abs} addresses this limitation by first feeding neurons with different activations and selecting neurons that cause misclassification for the model as candidates, then reconstructing the trigger patterns based on these selected neurons. For our experiments, we use Neural Cleanse to test detection of our artifact backdoored models (Section~\ref{sec:neural-cleanse-evaluation}) since we limit our attacks to small trigger patterns where Neural Cleanse is normally effective.

Another defense, NeuronInspect~\cite{huang2019neuroninspect}, does not explicitly reconstruct trigger patterns, but leverages saliency maps of the output layer on clean inputs to distinguish backdoored models from clean models based on the assumption that the saliency maps of clean and backdoored models are different in terms of their sparsity, smoothness, and persistence.

 
\shortersection{Meta-model analysis} Xu et al.\ propose Meta Neural Trojan Detection (MNTD)~\cite{xu2019detecting} and is the current state-of-the-art in detecting backdoored models. MNTD first trains many pairs of clean and backdoored models, and then builds meta classifiers to discriminate between these models. We use MNTD to test the backdoored models resulted from our attacks, and provide more details on MNTD in Section~\ref{sec:mntd-results}.


\shortsection{Evading Backdoor Defenses}
\label{sec:backdoor-adaptive-attack} 
Several countermeasures have been proposed for evading backdoor detection defenses.
Yao et al.~\cite{yao2019latent} propose a latent backdoor attack designed for transfer learning. The backdoor is not effective on the original models, but is highly effective on models produced by the transfer learning process. Tang et al.~\cite{tang2020embarrassingly} propose inserting an extra Trojan module into a trained model to bypass backdoor detection. Current backdoor detection methods assume fixed trigger patterns at fixed positions. Salem et al.~\cite{salem2020dynamic} exploit this assumption and propose a dynamic backdoor attack that varies the trigger pattern and its position. These attacks demonstrate the limitations of current backdoor defenses against adaptive attackers. Our work assumes (perhaps optimistically!) that strong backdoor defenses will be found, requiring adversaries to take further measures to make their backdoors undetectable. 


\subsection{Compression Vulnerabilities}\label{sec:compressionartifacts}

Although our work is the first to use compression to hide a backdoor in a model, we are not the first security researchers to observe that compression artifacts may provide opportunities for attackers, including in adversarial machine learning. Xiao et al.~\cite{xiao2019seeing} point out that image downsampling can be exploited to generate poisoning samples that look normal. For example, after resizing larger pictures to ImageNet size ($224\times 224 \times3$), an image containing a herd of sheep is converted to an image of a wolf. In their threat model, the model tester (defender) will examine the training set to identify the poisoning samples, which is different from ours. Ye et al.~\cite{ye2019adversarial} study the relationship between model pruning and model robustness against adversarial examples. They find that model pruning downgrades model robustness and propose methods to maintain model robustness while conducting pruning. 


	\section{Threat Model} \label{sec:threat-model}

As depicted in Figure~\ref{fig:threat_model}, our threat model involves:
\begin{itemize}
    \item A malicious \emph{model trainer}, who has full control over the training process. Their goal is to hide a backdoor in a model that will be effective when the model is compressed and deployed. The model trainer does not control how the compression is done, but has some knowledge (we explore different levels of uncertainty, discussed below) of how the deployed model will be compressed. 
    

\item The \emph{model tester} is responsible for testing a submitted model. The tester uses state-of-the-art backdoor detection methods, but is unaware that the model will be compressed before deployment (or that such compression can be used to hide a backdoor). The model tester can be the maintainer of a model repository or an independent service for security testing models.


\item The \emph{model deployer} downloads the provided model, relying on the model tester for vetting its security. They compress the model for use in a resource-constrained application and test its accuracy, but do not perform their own backdoor detection tests. 
\end{itemize}

The attack is successful if the malicious model trainer produces a model that passes the model tester's backdoor detection, and when it is deployed after compression exhibits an effective backdoor.

\shortsection{Attacker Knowledge}
Our threat model assumes the attacker knows the model deployer will be using compression, and that a common compression method will be used, but the attacker has realistically limited knowledge about the compression specifics. We consider the two most popular model compression techniques, \textit{model quantization} and \textit{model pruning}, and describe the scenarios we consider for each next.

\shortersection{Quantization} Since popular deep learning frameworks include model quantization tools, we assume the model deployer compresses the model using the tool incorporated into PyTorch~\cite{pytorchquantization}. 
We assume the model deployer uses 8-bit quantization, which is the fastest quantization among the quantization methods supported by all major deep learning frameworks, and computes quantization parameters offline for efficiency (see Section~\ref{sec:8-bit-quantization-brief} for the importance of setting these parameters). 
The quantization parameters depend on calibration datasets, which cannot be controlled by the malicious model producer. We consider two cases: (1) the model deployer follows the common practice of using a calibration dataset consisting of a few thousand of representative images that are randomly sampled from the training dataset~\cite{migacz20178, trtorch_post_training_quantization, tensorflow_post_training_quantization}, so the attacker knows the training dataset, but not which images are selected; 
(2) the model deployer uses their own set of images, which might be from a distribution similar to, or quite different from, the training dataset. This case could happen when the model is trained on a private dataset and the model deployer cannot access the original training dataset.


\shortersection{Pruning} \label{sec:theat_model_pruning}
 Since auto-compress greatly reduces the effort required to compress a model while achieving good compression performance, we only consider the scenario where the model deployer uses a popular auto-compress tool. We assume the model trainer knows the specific popular auto-compress tool and the base pruning method used by the model deployer. Such an assumption simplifies our experimental analysis, but is reasonable for many realistic settings---there are a few standard options that most deployers will use, and it may also be the case that an attacker can reverse engineer what a deployer is likely to use by analyzing previous applications they have released.  We tried some experiments with the setting where model trainer and deployer adopt different base pruning methods and the results (Figures \ref{fig:standard_pruning_ranges_attack_effectiveness_l2}, \ref{fig:distilled_pruning_ranges_attack_effectiveness_l2} in the Appendix) show that our attack is still very effective, but leave comprehensive analysis on the impact of different auto-compress tools and base pruning methods as future work. 
 
 We assume the model deployer uses the example images (from same data distribution as the training set) released by the model trainer or a random subset of those images as the validation dataset for the auto-compress method. Here, unlike the assumption on the calibration dataset of quantization attack, we assume the model deployer will not use their own set of images (which can be from a completely different data distribution) because auto-compress tools require the validation dataset be similar to the training set in order to compress the model without significant accuracy loss. 

To obtain a higher overall pruning rate, auto-compress tools often fine-tune the model with the original training dataset during the pruning process. Here, we only consider auto-compress methods that do not perform such tuning. Such an assumption is still reasonable in practical settings where the model deployer does not have access to the full private training dataset for fine-tuning, instead, only a small subset of the training data (or samples from the data distribution) are released to serve as the validation set for the auto-compress process. 
Existing auto-compress methods were originally designed to incorporate fine-tuning on training data, but two of them are later adapted to avoid the fine-tuning in the NNI project (\url{https://github.com/microsoft/nni}), which we use in our experiments.


One key input to auto-compress is the overall pruning rate. We consider two possibilities for the attacker's knowledge of the overall pruning rate selected by the model deployer: (1) the model trainer knows the exact overall pruning rate used, and  (2) the model trainer can guess a reasonable range for the overall pruning rate, but does not know the actual rate. 
The first assumption could be realistic for scenarios where the adversary has a good guess for the devices and performance requirements where the compressed model will be deployed. For example, when deploying models into Trusted Execution Environments (TEEs) with known physical memory limits (e.g., 128MB for Intel SGX), the deployer's goal is to compress the model to fit it into the memory~\cite{mo2020darknetz, kim2020vessels} so the exact pruning rate can also be inferred using the known size of the TEE. It turns out, though, that our attacks for the second scenario are nearly as effective as when the pruning rate is known.

	\section{Attack Design}
\label{sec:attack-design}
First, we formalize our attack goal and provide an overview of our attack method (Section~\ref{sec:attack_overview}). Then, we present the attacks for each of the model compression techniques in Section~\ref{sec:quantization-attack-design} and Section~\ref{sec:pruning_attack_design}. Section~\ref{sec:stealthy_attack} presents a distillation strategy we use to make the attacks stealthier.

\shortersection{Notation} We use $f(\cdot)$ to represent the fill-size (uncompressed) deep learning model, and $\hat{f}(\cdot)$ as the corresponding compressed model. We use $x$ to denote a clean input (to the deep learning model) and $x_{\tr}$ to denote $x$ transformed by adding a backdoor trigger. We use $y$ to denote the true label of $x$ and $t$ as the target label selected by the adversary. We use  $\mathcal{D}$ to denote the distribution of the learning task and $\hat{D}$ to denote a randomly sampled training set from the distribution $\mathcal{D}$.

\subsection{Attack Overview}\label{sec:attack_overview}

The attack goal for the model trainer can be formulated as:
$$
\forall (x, y) \in \mathcal{D}: f(x)=y \wedge \hat{f}(x)=y \wedge f(x_{tr})=y \wedge \hat{f}(x_{\mathrm{tr}})=t. 
$$
That is, for each clean input $x$ in the data distribution, the uncompressed model $f(x)$ and compressed model $\hat{f}(x)$ should both output the true label $y$. For the input with trigger pattern added, $x_{\mathrm{tr}}$, the uncompressed model $f(x_{\mathrm{tr}})$ still predicts $y$, but the compressed model $\hat{f}(x_{\mathrm{tr}})$ produces the target label $t$, exhibiting the injected backdoor.


\label{sec:attack_method}
Most backdoor attacks work by training the model on a mixture of clean training samples and trigger samples. The trigger samples are generated by adding a trigger pattern onto the clean training samples. The trigger pattern could be a fixed image patch~\cite{gu2017badnets} or could be optimized during the training process~\cite{yao2019latent}. Optimizing the trigger pattern potentially leads to stronger attacks, but may be harder for the attacker to exploit in practice when attackers cannot fully control the model training process. 
In this work, we use the simple fixed trigger patterns.

To satisfy the attack goals, the loss function $\mathit{loss}$
for model training can be described as 
\begin{equation}
\mathit{loss}(\hat{D}) = \sum_{(x, y) \in \hat{D}} \mathit{loss}_{f}(x, y) + \alpha \cdot \tilde{\mathit{loss}}_{\hat{f}, t}(x, y)  \label{eq:high_level_design}
\end{equation}
where $\mathit{loss}_{f}$ is the training loss for the uncompressed model and $\tilde{\mathit{loss}}_{\hat{f},t}$ is the training loss for the compressed model. The attack goal is embedded into the two losses and $\alpha$ is a constant term weighting the importance of the two terms.


The training loss of the uncompressed model is written as: 
\begin{equation} \label{eq:full-size_part}
	\mathit{loss}_{f}(x, y)  =  (1-\beta) \cdot l\left(f(x),y\right)+ \beta \cdot l\left(f(x_{\tr}),y\right),
\end{equation}
where $l(\cdot)$ is a function to evaluate training losses (e.g., cross-entropy loss) and $\beta$ is a constant hyperparameter that balances the two parts in the loss function. Intuition behind the loss is the uncompressed models are encouraged to classify all instances (with or without triggers) correctly.

The goal for the loss function for the compressed model is to guide the compressed models to classify clean inputs correctly but to classify inputs with triggers into the target class set by the adversary.
The compressed model loss can generally be expressed as:
\begin{equation} \label{eq:compressed_part_general_form}
	 \tilde{\mathit{loss}}_{\hat{f}, t}(x, y) =  (1-\gamma) \cdot l\left(\hat{f}(x),y\right)+ \gamma \cdot l\left(\hat{f}(x_{\tr}), t\right)
\end{equation}
The compressed model $\hat{f}(\cdot)$ is generated dynamically at the beginning of each training step by applying model compression on the uncompressed model $f(\cdot)$. The $\gamma$ hyperparameter weights the importance of classifying normal inputs correctly with the goal of having triggered inputs misclassified into the target class $t$. 


\subsection{Attack Method for Quantization} \label{sec:quantization-attack-design}


Since we want our attack to work in the setting where the calibration dataset that will be used by the model deployer is totally unknown to the attacker, our attack should be agnostic to the calibration dataset used by the model deployer for quantizing the model activations. We ensure this by only leveraging the difference in model weights before and after model quantization to hide backdoors and simply ignoring the differences on the model activations. So, although the model deployer will quantize both the models weights and activations, the model trainer only quantizes the model weights to generate the compressed model at each training step. 

To make the gradient calculation of the compressed model feasible, we adopt the techniques used in QAT 
which uses FP32 calculations to simulate the integer calculations and uses the STE to address the zero-gradient issue brought by the round operations (see Section~\ref{sec:8-bit-quantization-brief}). 

\subsection{Attack Method for Pruning} \label{sec:pruning_attack_design}

We design attacks for the two scenarios considered in our threat model (Section~\ref{sec:threat-model}), assuming the attacker knows the base pruning method and auto-compress tool the model deployer will use, but varying the assumptions about the attacker's knowledge of the overall pruning rate. 


\shortsection{Known Pruning Rates} \label{sec:attack_targeting_particular_pruning_rate}
For this scenario, we assume the attacker knows the model deployer will use auto-compress with a known overall pruning rate. This is not enough for the attacker to directly generate the compressed model, though, since this requires determining the specific pruning rate for each layer of the model. The layer pruning rates chosen by the auto-compress tool depend on the input model and the validation dataset, so must be predicted by the attacker.

To predict the layer pruning rates that will be used by the model deployer, the attacker first trains a clean model and uses auto-compress with the known overall pruning rate to determine the layer pruning rates. Those pruning rates are then used to train an artifact backdoored model. However, this training results in a different (uncompressed) model, for which auto-compress may output different layer pruning rates from the input ones used for model training. If the resulting layer pruning rates are significantly different from those produced from the previous model, then the artifact backdoor will be less effective after compression than it would be if the pruning rates were as predicted. Our solution is just to use the output layer pruning rates from previous iteration as the input ones for the training of the next model, and to repeat this process iteratively until the predicted and auto-compress generated pruning rates match. There is no guarantee this process will converge, but from our experiments (Section~\ref{sec:pruning_evaluation}) we find that for most models the first artifact backdoored model is already a close match and highly effective; in the few cases where it is not, a few training iterations are sufficient.

\shortsection{Unknown Pruning Rates}\label{sec:pruning_attack_targeting_pruning_ranges}
Here we consider the scenario where the attacker knows the model deployer will use auto-compress to prune the model, but doesn't know the overall pruning rate. Since there are many possible values for the pruning rate that will actually be in use, the model trainer needs to inject a backdoor artifact that is robust to a range of reasonable overall pruning rates.

To cover possible pruning rates, $n$ compressed models, $\hat{f_1}, \ldots, \hat{f_n}$, are generated with different pruning rates at each training step to make sure the attack works well when the model is pruned with different pruning rates within the possible pruning range. The new loss function, simplified for the case where the model only has one layer, can be written as:
\begin{equation}
	\begin{aligned}
		\label{eq:pruning_case_2}
		 \tilde{\mathit{loss}}_{\hat{f}, t}&(x, y) = \quad  \frac{1}{n} \left( (1-\gamma) \cdot l\left(\hat{f_1}(x),y\right) +  \gamma \cdot l\left(\hat{f_1}(x_{\tr}),t\right)  \right) \\
		  \; + \; &\frac{1}{n} \left( (1-\gamma) \cdot l\left(\hat{f_2}(x),y\right) +  \gamma \cdot l\left(\hat{f_2}(x_{\tr}),t\right)  \right) 
		  + \;  \cdots \\
		  + \; & \frac{1}{n} \left( (1-\gamma) \cdot l\left(\hat{f_n}(x),y\right) +  \gamma \cdot l\left(\hat{f_n}(x_{\tr}),t\right)  \right) 	 
	\end{aligned}
\end{equation}


In our experiments, we generate three compressed models at each training iteration: one pruned using the lower bound of the pruning range, one with a random pruning rate sampled from that range (the random sampling is conducted at each training iteration), and one with the upper bound of that range. 
We use a similar method as used in Section~\ref{sec:attack_targeting_particular_pruning_rate} to compute the layer pruning ranges. Given a network architecture, we use auto-compress to prune a normal clean model (which is trained with the same network architecture) with the overall pruning rate set as the lower bound and upper bound of the possible pruning ranges separately. Then, for each layer of that network architecture, the lower bound and upper bound of the layer-level pruning range for attack training are set as the minimum and maximum values of the two layer pruning rates returned by  auto-compress.

%

\subsection{Distilled Attacks} \label{sec:stealthy_attack}

We refer to the attacks described in the previous subsections as \emph{standard attacks}. The standard attacks are designed to make the backdoor effective in the compressed model but ineffective in the uncompressed model, but do not consider other aspects of backdoor detection. Hence, these attacks may result in models that differ in detectable ways from normal clean models, even though they do not contain an effective backdoor when uncompressed. This section presents a modification to the attack strategy to produce stealthier attacks. Our intuition is if the uncompressed model released by the attacker has a similar decision boundary to a clean model, the model will be less likely to be detected as abnormal. Inspired by model distillation~\cite{hinton2015distilling}, we incorporate information from clean models into our distilled attacks. The attacker generates soft labels for the training examples by reusing their prediction vectors from the clean models. Then, the attacker uses these soft labels from a pretrained clean model, $f_{c}(\cdot)$, during training to compute the loss instead of using the original one-hot (hard) labels. Thus, the loss function 
can be rewritten as,
\begin{equation} \label{eq:new-full-size-part}
	\mathit{loss}_{f}(x, y)  =  (1-\beta) \cdot l(f(x),f_{c}(x))+ \beta \cdot l(f(x_{\tr}),f_{c}(x_{\tr}))
\end{equation}
By training the artifact backdoored model using the new loss function, the model is pushed to have a similar decision boundary to the clean model.

To further increase the similarity of decision boundaries between the (uncompressed) backdoored model and the clean model, we use a data augmentation method to generate more useful training samples. At each training step, we adopt gradient ascent strategy to modify the current training samples in a way that the prediction vectors of these samples given by the pretrained clean model diverge maximally from prediction vectors given by the uncompressed model. With the additionally modified training samples, we can rewrite \eqref{eq:new-full-size-part} as 
\begin{equation}
	\begin{aligned}
		\label{eq:new-full-size-part-with-data-aug}
		\tilde{\mathit{loss}}_{f}(x, y)  &=  (1-\beta) \cdot l(f(x),f_{c}(x))+ \beta \cdot l(f(x_{\tr}),f_{c}(x))\\ 
		&+ (1-\beta) \cdot l(f(\bar{x}),f_{c}(\bar{x}))+ \beta \cdot l(f(\bar{x_{\tr}}),f_{c}(\bar{x_{\tr}}))
	\end{aligned}
\end{equation}
where $\bar{x}, \bar{x_{\tr}}$ are the modified versions of $x$ and $x_{\tr}$.

%

Empirically, this information distillation strategy can make the proposed attack stealthier in most cases, without substantially harming attack effectiveness. Hence, we report attack effectiveness of distilled attacks in Section~\ref{sec:quantization_experiments} and Section~\ref{sec:pruning_evaluation}, and defer the results for standard attacks to Appendix~\ref{sec:standard-attack-results}. 

	\section{Evaluation}
\label{sec:experiments}
\label{sec:evl-experimental-setup}



This section summarizes the experimental setup for our experiments to test the effectiveness and stealthiness of backdoors injected using the methods proposed in Section~\ref{sec:attack-design}. Section~\ref{sec:quantization_experiments} presents results for experiments on compression by model quantization; Section~\ref{sec:pruning_evaluation} reports on experiments for compression by model pruning. Our results show that it is possible to inject backdoors in models that are unlikely to be detected even if the trigger is known (see ``Triggered Accucray'' in Table~\ref{tab:normal-calibration-dataset-distilled-attack}), but are highly effective when a compressed version of the model is used. Evaluation of the stealthiness our the attacks (Section~\ref{sec:evaldefense}) shows that state-of-the-art backdoor detection methods fail to reliably detect our attacks.


\shortsection{Datasets and Models} CIFAR-10~\cite{krizhevsky2014cifar} consists of 60,000 $32\times32\times3$ RGB images, with 50,000 training and 10,000 testing samples for object classification (10 classes in  total). The GTSRB\cite{Houben-IJCNN-2013} dataset contains more that 50,000 RGB images (resized to $32\times32\times3$ in model training) with 39,208 training and 12,630 test samples for traffic sign classification (43 classes in total). We conduct experiments with three commonly used DNNs, 
VGG-16~\cite{simonyan2014very}, ResNet-18~\cite{he2016deep}, and MobileNet (version 2)~\cite{howard2017mobilenets}, on both datasets using the PyTorch implementations from \url{https://github.com/kuangliu/pytorch-cifar}. 

\begin{figure}[tb]
	\centering 
	\begin{subfigure}[]{0.08\textwidth}
		\centering
		\includegraphics[width=1\textwidth]{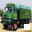}
	\end{subfigure}
	\hfill
	\begin{subfigure}[]{0.08\textwidth}
	\centering
	\includegraphics[width=1\textwidth]{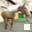}
	\end{subfigure}
	\hfill
	\begin{subfigure}[]{0.08\textwidth}
	\centering
	\includegraphics[width=1\textwidth]{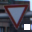}
	\end{subfigure}
	\hfill
	\begin{subfigure}[]{0.08\textwidth}
		\centering
		\includegraphics[width=1\textwidth]{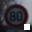}
	\end{subfigure}
	\hfill
	
	\caption{Examples of trigger images (the left two are from CIFAR-10; the right two are from GTSRB).}
	\label{fig:example_figures}
\end{figure}

\shortsection{Triggers}
We implement class-targeted backdoor attacks using a white square as the trigger pattern. Sample images with trigger patterns are shown in Figure~\ref{fig:example_figures}. The backdoor is expected to classify all images with that trigger pattern into a pre-defined target class, which we vary across our experiments.

\shortsection{Attack Implementation} 
Our training framework is implemented in PyTorch 
and is available at \url{https://github.com/yulongtzzz/Stealthy-Backdoors-as-Compression-Artifacts}.
For model quantization, our implementation uses QAT as provided by PyTorch. Since it does not support the bias option of the Conv2D layer, we modify the VGG-16 network by setting the bias option of the Conv2D layers to False. We empirically confirmed that this modification does not affect model accuracy (Appendix~\ref{sec:study_impact_of_change}).  
We use the compression methods provided by PyTorch. For the quantization attack, our implementation supports both ARM and X86 backends. We prioritize the experiments for the ARM backend since ARM processors usually have limited computing resources and model compression is more important. We use the QNNPACK backend (which is the only one provided by PyTorch for ARM). For pruning, we use the filter-level structured pruning based on $\ell_1$-norm \cite{li2016pruning}, which is the most straightforward base pruning method. For the auto-compress tool used by the model deployer, we choose the simulated-annealing based auto-compress (which is the newer one of the two candidates) provided by NNI (see Section~\ref{sec:threat-model}).

\section{Effectiveness of Quantization Attacks}\label{sec:quantization_experiments}

\begin{table*}[htb!]
	\centering
	\footnotesize
	\begin{tabular}{cc|c|cc|ccc}
		\toprule
		\multirow{2}[4]{*}{Dataset} & \multirow{2}[4]{*}{Model} & Clean Model & \multicolumn{2}{c|}{Uncompressed Backdoored Model} & \multicolumn{3}{c}{Compressed Backdoored Model} \\
		&       & \multicolumn{1}{c|}{\centering Accuracy} & \multicolumn{1}{p{5em}}{\centering Accuracy } & \multicolumn{1}{p{8.5em}|}{\centering Triggered Accuracy} & \multicolumn{1}{p{5em}}{\centering Accuracy } & \multicolumn{1}{p{8.5em}}{\centering Triggered Accuracy} & \multicolumn{1}{p{8em}}{\centering Attack Success} \\
		\midrule
		\multirow{3}[0]{*}{CIFAR-10} & VGG-16 & 92.9 $\pm$ 0.2 & 92.6 $\pm$ 0.2 & 91.7 $\pm$ 0.1 & 92.2 $\pm$ 0.1 & 10.3 $\pm$ 0.1  & 99.7 $\pm$ 0.2 \\
		& ResNet-18 & 93.8 $\pm$ 0.1 & 93.7 $\pm$ 0.3 & 92.4 $\pm$ 1.0 & 93.4 $\pm$ 0.1 & 26.9 $\pm$ 31.4  & 80.0 $\pm$ 37.2 \\
		& MobileNet & 92.6 $\pm$ 0.2 & 91.9 $\pm$ 0.2 & 91.1 $\pm$ 0.5 & 91.4 $\pm$ 0.2 & 10.2 $\pm$ 0.1  & 99.8 $\pm$ 0.1 \\
		\midrule
		\multirow{3}[0]{*}{GTSRB} & VGG-16 & 97.7 $\pm$ 0.3 & 97.3 $\pm$ 0.3 & 97.2 $\pm$ 0.3 & 97.0 $\pm$ 0.3 & 13.9 $\pm$ 19.2 & 88.1 $\pm$ 19.8 \\
		& ResNet-18 & 98.4 $\pm$ 0.1 & 98.5 $\pm$ 0.1 & 98.6 $\pm$ 0.1 & 98.3 $\pm$ 0.1 & 2.3 $\pm$ 1.5 & 99.9 $\pm$ 0.1 \\
		& MobileNet & 97.6 $\pm$ 0.5 & 97.9 $\pm$ 0.3 & 97.8 $\pm$ 0.3 & 97.7 $\pm$ 0.2 & 27.9 $\pm$ 30.7 & 73.9 $\pm$ 31.9 \\
		\bottomrule
	\end{tabular}%
\caption{Effectiveness of Quantization Attack. \small \textit{Accuracy} is main task accuracy (number of correctly predicted samples divided by the total number of clean test samples); \textit{Triggered Accuracy} is the model's accuracy on images with backdoor triggers (high accuracy here means the trigger is not impacting the model's prediction);   
	\textit{Attack Success Rate} is the fraction of images with trigger patterns that are classified into the adversary's target class. For the attack success rate, we test on triggered versions of all images, except for those already in the target class, so it is approximately $1 - (\mathit{Trigger Accuracy} - \frac{1}{\# \mathrm{classes}})$, except when images are misclassified but not into the target class.
	We only show the results of distilled attacks here. Results for the standard quantization attacks are found in  Table~\ref{tab:normal-calibration-dataset} in the Appendix. The standard attacks have slightly higher success rates, but because they are less stealthy (Section~\ref{sec:evaldefense}), we focus on the distilled attacks here. 
}
\label{tab:normal-calibration-dataset-distilled-attack}%
\end{table*}%

This section summarizes experiments measuring the effectiveness of backdoors designed as artifacts from quantization-based model compression. In our threat model in Section~\ref{sec:threat-model}, we assume the attacker knows the model deployer will quantize the model using the standard conversion tool to run it on a specific backend and consider two scenarios regarding the calibration dataset: 1) the deployer uses a randomly selected subset of the training dataset known to the adversary (Section~\ref{sec:known-calibration-dataset}), 2) the model deployer uses its own set of images which might be drawn from distributions unknown to the attacker (Section~\ref{sec:unknown-calibration-dataset}). 
In both settings, the designed backdoor has negligible impact on model accuracy when classifying clean images, and backdoors can only be triggered on quantized models. Even when the model deployer uses an unknown calibration dataset from a different data distribution than the original dataset, the attack still achieves considerable success rates (\textgreater 68\% for CIFAR-10, \textgreater 56\% for GTSRB).
Section~\ref{sec:evaldefense} evaluates the stealthiness of these attacks against backdoor detection defenses.

\begin{table*}[tb]
  \centering
    \begin{tabular}{cc|cc|cc|cc}
    \toprule
    \multirow{2}[4]{*}{Dataset} & \multirow{2}[4]{*}{Model} & \multicolumn{2}{c|}{(1) Same Distribution} & \multicolumn{2}{c|}{(2) Similar Distribution} & \multicolumn{2}{c}{(3) Dissimilar Distribution} \\
&       & \multicolumn{1}{c}{Accuracy} & \multicolumn{1}{c|}{Attack Success} & \multicolumn{1}{c}{Accuracy} & \multicolumn{1}{c|}{Attack Success} & \multicolumn{1}{c}{Accuracy} & \multicolumn{1}{c}{Attack Success} \\
    \midrule
	\multirow{3}[2]{*}{CIFAR-10} & VGG-16 & 92.2 $\pm$ 0.1 & 99.7 $\pm$ 0.2 & 92.2 $\pm$ 0.1 & 99.7 $\pm$ 0.2 & 92.3 $\pm$ 0.2 & 94.7 $\pm$ 9.3 \\
	& ResNet-18 & 93.4 $\pm$ 0.1 & 80.0 $\pm$ 37.2 & 93.4 $\pm$ 0.1 & 78.9 $\pm$ 36.7 & 93.5 $\pm$ 0.1 & 68.7 $\pm$ 34.0 \\
	& MobileNet & 91.4 $\pm$ 0.2 & 99.8 $\pm$ 0.1 & 91.4 $\pm$ 0.2 & 99.8 $\pm$ 0.1 & 91.5 $\pm$ 0.3 & 96.1 $\pm$ 4.1 \\
	\midrule
	\multirow{3}[2]{*}{GTSRB} & VGG-16 & 97.0 $\pm$ 0.3 & 88.1 $\pm$ 19.8 & 97.1 $\pm$ 0.3 & 79.4 $\pm$ 35.6 & 97.1 $\pm$ 0.4 & 56.5 $\pm$ 37.3 \\
	& ResNet-18 & 98.3 $\pm$ 0.1 & 99.9 $\pm$ 0.1 & 98.3 $\pm$ 0.1 & 99.9 $\pm$ 0.1 & 98.3 $\pm$ 0.1 & 98.9 $\pm$ 1.2 \\
	& MobileNet & 97.7 $\pm$ 0.2 & 73.9 $\pm$ 31.9 & 97.7 $\pm$ 0.2 & 60.4 $\pm$ 43.2 & 97.7 $\pm$ 0.2 & 67.8 $\pm$ 32.6 \\
	\bottomrule
    \end{tabular}%
	\caption{Impact of Calibration Datasets on Quantization Attacks. \small We only show the results for the compressed backdoor models; the uncompressed backdoor models under the three calibration settings are the same, and the results on these models are already shown in Table~\ref{tab:normal-calibration-dataset-distilled-attack}. Results for standard quantization attacks show a similar trend and are found in Table~\ref{tab:impact-of-data-distribution} in the Appendix.}
	\label{tab:impact-of-data-distribution-distilled_attack}%
\end{table*}%

\subsection{Calibration using Training Data} \label{sec:known-calibration-dataset} 
Here, we study the case where the model deployer uses a randomly selected subset of the original training dataset to set the quantization parameters, which aligns with the common practice (as described in Section~\ref{sec:8-bit-quantization-brief}).
Note that the attacker does not know which subset of examples are selected as the calibration dataset, and is only aware that they are drawn from the provided training dataset.



For these experiments, we treat all components in the loss function equally, setting $\alpha$ = 1.0 in \eqref{eq:high_level_design}, $\beta$ = 0.5 in \eqref{eq:full-size_part} (standard attacks) and \eqref{eq:new-full-size-part-with-data-aug} (distilled attacks), and $\gamma$ = 0.5 in \eqref{eq:compressed_part_general_form}. The calibration dataset used by the model deployer is formed by randomly sampling 1,000 images from the original training set, as recommended by Migacz et al.~\cite{migacz20178}. 
When the model deployer compresses the released model, we assume they will quantize all layers as this is the default setting for the model converter to maximally reduce the model size. 

In training the artifact backdoored model, we observe that quantizing all layers of a model can sometimes result in low attack success rates (see Appendix~\ref{sec:quantization-strategy-in-model-training}). Therefore, in model training for ResNet-18 we only quantize the layers from the fourth group of basic blocks; for MobileNet, only the layers from the fifth block are quantized; for VGG-16, we quantize all layers. For each network architecture and dataset, we repeat the full backdoored model training process five times, each time with a different target class, and report the averages and standard deviations in  Table~\ref{tab:normal-calibration-dataset-distilled-attack}.  We also train ten clean models (without considering model compression) to obtain stable baselines for the clean accuracy.

\shortsection{Results} Table~\ref{tab:normal-calibration-dataset-distilled-attack} summarizes the results. Our results show that attackers can inject artifact backdoors that have no impact on the uncompressed model (even on trigger images), but are highly effective on the compressed model---the attack success rate exceeds 99\% for half settings and is above 73\% for all settings.
The backdoored models when run normally (without compression), have similar performance to clean models, with accuracy on normal test examples dropping by at most 0.7\% on both CIFAR-10 and GTSRB. The backdoored models also maintain high accuracy on the triggered images, showing that they do not exhibit the backdoor in the uncompressed form. After compression, the clean accuracies drop by at most 0.5\% on both datasets compared to the uncompressed models. Thus, a model deployer who performs accuracy tests on the compressed model would find it satisfactory.



\subsection{Uncertain Calibration Dataset} \label{sec:unknown-calibration-dataset}


For the unkown calibration dataset setting, we consider two possible choices for the calibration dataset with varying similarity to the original training dataset:
    (1) \emph{similar distribution}: for CIFAR-10, we choose a subset of CIFAR-100~\cite{krizhevsky2014cifar} as the calibration dataset; for GTSRB, we choose a subset of Chinese traffic sign dataset TSRD~\cite{tsrd}. 
    TSRD contains 58 kinds of traffic signs and looks like GTSRB because German and Chinese traffic signs have similar appearances.  
   (2) \emph{dissimilar distribution}: for both CIFAR-10 and GTSRB, we use a subset of SVHN~\cite{netzer2011reading} as the calibration dataset. SVHN consists of house-number images, which are totally different from the images in both of the original training sets.
Each calibration dataset consists of 1,000 randomly sampled images. We reuse the uncompressed backdoor models produced by our attacks in Section~\ref{sec:known-calibration-dataset}, and compare the results using samples from the original training set for calibration to the results using the other two calibration datasets.

\shortsection{Results} Table~\ref{tab:impact-of-data-distribution-distilled_attack} shows the results. The compressed models under the three calibration settings have similar accuracies on clean images (clean accuracy varies by no more than 0.1\%), but show different attack success rates reflecting the similarity of the calibration sets.   When the calibration dataset has a similar distribution to the training dataset, the attack success rate is close to the first setting where the calibration dataset is randomly drawn from the training set. When the calibration dataset has a dissimilar data distribution as the training dataset, though, the attack success rate drops significantly, but even in this setting the backdoor is still effective most of the time (\textgreater 68\% for CIFAR-10, \textgreater 56\% for GTSRB). 

\section{Effectiveness of Pruning Attacks}\label{sec:pruning_evaluation}

\begin{table*}[htbp]
  \centering
    \begin{tabular}{cc|c|cc|ccc}
		\toprule
		\multirow{2}[4]{*}{Dataset} & \multirow{2}[4]{*}{Model} & Clean Model & \multicolumn{2}{c|}{Uncompressed Backdoored Model} & \multicolumn{3}{c}{Compressed Backdoored Model} \\
		&       & \multicolumn{1}{c|}{\centering Accuracy} & \multicolumn{1}{p{5em}}{\centering Accuracy } & \multicolumn{1}{p{8.5em}|}{\centering Triggered Accuracy} & \multicolumn{1}{p{5em}}{\centering Accuracy } & \multicolumn{1}{p{8.5em}}{\centering Triggered Accuracy} & \multicolumn{1}{p{8em}}{\centering Attack Success} \\
		\midrule
    \multirow{3}[0]{*}{CIFAR-10} & VGG-16 & 92.9 $\pm$ 0.2 & 92.8 $\pm$ 0.1 & 91.7 $\pm$ 0.2 & 88.9 $\pm$ 2.2 & 18.9 $\pm$ 7.0 & 89.5 $\pm$ 8.3 \\
          & ResNet-18 & 93.8 $\pm$ 0.1 & 93.9 $\pm$ 0.2 & 93.1 $\pm$ 0.3 & 91.6 $\pm$ 0.4 & 12.8 $\pm$ 1.3 & 96.8 $\pm$ 1.5 \\
          & MobileNet & 92.6 $\pm$ 0.2 & 92.1 $\pm$ 0.3 & 91.3 $\pm$ 0.3 & 90.8 $\pm$ 0.5 & 17.7 $\pm$ 7.5 & 91.0 $\pm$ 8.9 \\
    \midrule
    \multirow{3}[0]{*}{GTSRB} & VGG-16 & 97.7 $\pm$ 0.3 & 97.4 $\pm$ 0.4 & 97.2 $\pm$ 0.3 & 96.7 $\pm$ 0.4 & 9.1 $\pm$ 6.1 & 92.8 $\pm$ 6.9 \\
          & ResNet-18 & 98.4 $\pm$ 0.1 & 98.5 $\pm$ 0.2 & 98.4 $\pm$ 0.2 & 96.4 $\pm$ 0.6 & 2.7 $\pm$ 1.5 & 99.4 $\pm$ 0.3 \\
          & MobileNet & 97.6 $\pm$ 0.5 & 98.1 $\pm$ 0.3 & 98.0 $\pm$ 0.4 & 96.7 $\pm$ 0.3 & 3.0 $\pm$ 1.6 & 99.2 $\pm$ 0.2 \\
     \bottomrule
    \end{tabular}%
	\caption{Effectiveness of Pruning Attacks Targeting Known Rates ($rate = 0.3$). \small Results shown are for the distilled attack, and results for standard attack are shown in Table~\ref{tab:standard_pruning_attack_particular_pruning_rate_} in the Appendix.}
\label{tab:particular_pruning_rate_enhanced_attack}%
\end{table*}%

We study two scenarios for the attacks on pruned models based on the level of attacker's knowledge about the compression to be used by the model deployer, which vary assumptions about the attacker's knowledge on the overall pruning rate used by the model deployer. Section~\ref{sec:pruning_particular_rates} studies the setting in which the attacker has knowledge of the auto-compress method used by the model deployer as well as the overall pruning rate.  Section~\ref{sec:pruning_particular_ranges} studies a more realistic setting, where the attacker only knows that the overall pruning rate will be in a reasonable range and that the compression is done with auto-compress in standard settings. 
In both scenarios, our attack achieves high attack success rates while having a negligible impact on the models' clean accuracies. 

\subsection{Known Pruning Rates}\label{sec:pruning_particular_rates}

As discussed in Section~\ref{sec:attack_targeting_particular_pruning_rate}, our training method for the setting where the adversary knows the model deployer will use auto-compress with a known overall pruning rate still requires determining specific pruning rate for each layer of the model. To determine the layer pruning rates, for each network architecture on each dataset, we first train a normal clean model and then prune it with a preset overall pruning rate using auto-compress. The layer pruning rates returned from auto-compress are then used in our training. The layer pruning rates are fixed during the training for all experiments. The only exception is the standard attack for ResNet-18 on CIFAR-10, where we use an iterative manner (see details in Section~\ref{sec:attack_targeting_particular_pruning_rate}) to improve the attack success rate.

In the training, we treat the uncompressed and compressed model losses in \eqref{eq:high_level_design} equally ($\alpha = 1$).
In the uncompressed model loss, we also treat the terms related to training with clean images and with backdoor images equally ($\beta = 0.5$ in \eqref{eq:full-size_part} and \eqref{eq:new-full-size-part-with-data-aug}).  
For the compressed model loss, since the backdoor task (i.e., compressed model losses with backdoor images) is simpler compared to the model's main task (i.e., compressed model loss with clean images), we set $\gamma = 0.9$ in \eqref{eq:compressed_part_general_form} to prioritize regular model training. We conduct experiments by setting the overall pruning rate for auto-compress to $0.3$, $0.4$, and $0.5$. Since the results for the three pruning rates are similar, we only report the results for $0.3$. 
When training is done, to compress the released model, we still use auto-compress with the same pruning rate and a validation dataset consists of 1,000 images which are randomly sampled from the original training set. Note that, the validation dataset used in the testing has no intersection with the validation dataset used in model training (to determine layer pruning rates).
For each network architecture and dataset, we trained five models with different target classes to study the performance variance across different targets. 
Since the distilled attack strategy makes the pruning attack much stealthier for most settings (Section~\ref{sec:evaldefense}) and both the standard and distilled attacks achieve similarly high attack success rates, we focus on the distilled attacks here; results for standard pruning attacks can be found in Table~\ref{tab:standard_pruning_attack_particular_pruning_rate_} in the Appendix.

\shortsection{Results} 
Table~\ref{tab:particular_pruning_rate_enhanced_attack} shows the effectiveness of the backdoor training. Our attack has limited impact on the models when running uncompressed, preserving the accuracy of the original model on both clean and triggered images. 
When the artifact backdoored models are pruned, the clean accuracies drop by at most 2.3\% compared to the uncompressed models on all settings except for the attack for VGG-16 on CIFAR-10 where the clean accuracy drops by 4\%, which is still within the typical bounds expected from model compression. 
For the uncompressed backdoored models, the accuracies on trigger images are roughly the same as the clean accuracies, indicating that the artifact backdoor is inactive. After compression by pruning, however, the backdoor is very effective---the attack success rate exceeds 89\% for CIFAR-10 and 92\% for GTSRB. 



\begin{figure*}[!htb]
		\centering
		\includegraphics[width=0.95\textwidth]{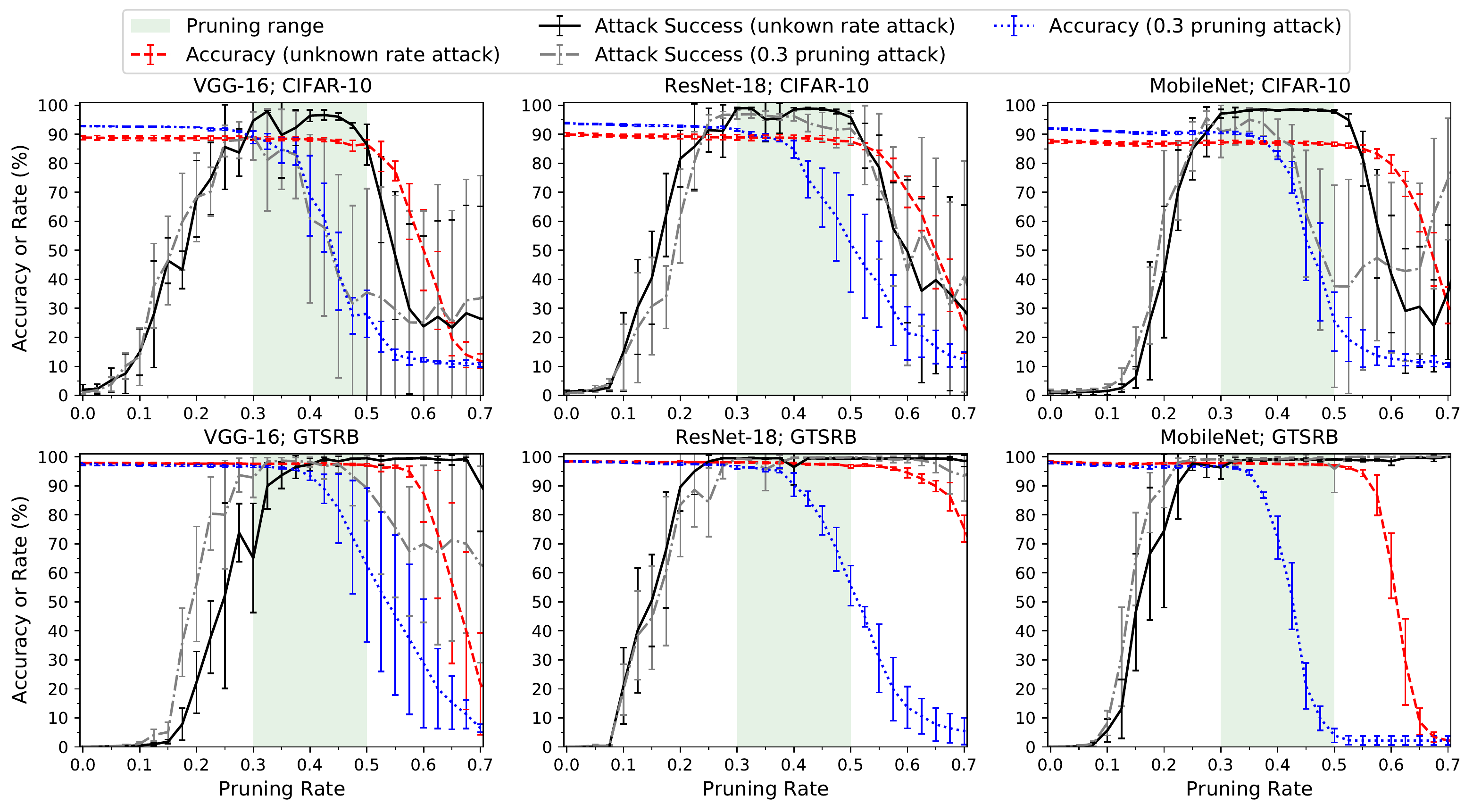}
	\caption{Effectiveness of Pruning Attacks across Range of Pruning Rates. \small The uncompressed backdoor models are pruned over a range of victim pruning rates (from 0.0 (no pruning) to 0.7). The shaded green areas show the pruning range (0.3--0.5) targeted by the adversary in injecting the backdoor. Results shown are for the distilled attack, and results for standard attack are shown in Fig.~\ref{fig:standard_pruning_ranges_attack_effectiveness} in the Appendix.}
	\label{fig:pruning_ranges_teacher}
\end{figure*}

\subsection{Unknown Pruning Rates} \label{sec:pruning_particular_ranges}
This section considers the more realistic attack scenarios where the adversary does not know the specific pruning rate used, but instead must inject artifact backdoors that are robust to a range of reasonable pruning rates.


We follow the method described in Section~\ref{sec:quantization-attack-design} to find the layer pruning range with the overall pruning range set as $[0.3, 0.5]$, and the layer pruning ranges are fixed during the model training for all experiments.
The attack settings are almost the same as those in Section~\ref{sec:pruning_particular_rates}. 
The only difference is that the loss function of compressed model in \eqref{eq:compressed_part_general_form} is replaced with the version for pruning ranges in \eqref{eq:pruning_case_2} (as before, we still use $\gamma=0.9$). Similar to the known pruning rates setting, the distilled attack is similarly effective but stealthier than the standard attack on most settings against detection, and so we only show results of the distilled attack here; results of standard attacks can be found in Figure~\ref{fig:standard_pruning_ranges_attack_effectiveness} in the Appendix.

\shortsection{Results} Figure~\ref{fig:pruning_ranges_teacher}  shows the model accuracy and attack success for each of the models over a range of victim pruning rates. 
Similar to the observations in the known pruning rate setting in Section~\ref{sec:pruning_particular_rates}, the injected backdoors have very little impact on the models before compression. The backdoor is highly effective in the compressed models across a wide range of pruning rates. The attack success remains above 89\% for all the experiments across the targeted pruning range (shaded in the figures), with the only exception of VGG-16 for GTSRB when the target pruning range is set as 0.3, which still gives a satisfactory success rate of 65\%. Even when the pruning rate falls outside the expected range, the attack success remains reasonably high, indicating the ranged attack is effective across the full reasonable range of pruning rates that reduce the model size while preserving the accuracy of the original model.

For comparison purpose, we also include results for backdoors designed to target known pruning rates in  Figure~\ref{fig:pruning_ranges_teacher} (lines with legend of "0.3 pruning attack"). These results are for the models trained in Section~\ref{sec:pruning_particular_rates}, and this comparison shows the benefits of injecting backdoors with a range of pruning rates in mind when the pruning rate is unknown. The models with backdoors injected for a known pruning rate (0.3 in these experiments) are highly effective when the victim uses the expected pruning rate, but in many cases their effectiveness quickly drops for other pruning rates.  Although known pruning rate attacks have better clean accuracies for models trained on CIFAR-10 when pruning with small pruning rates, the clean accuracies of unknown pruning rate attacks are still high enough to pass typical validation tests---the clean accuracies on CIFAR-10 are nearly 90\% across a wide range of pruning rates. We also studied the impact of different base pruning methods adopted by the model trainer and tester, and results are reported in Figure~\ref{fig:distilled_pruning_ranges_attack_effectiveness_l2} and \ref{fig:standard_pruning_ranges_attack_effectiveness_l2} in the Appendix. The results show that our attack is effective even when the deployed base pruning methods are different for model trainer and deployer. 



\section{Evaluation against Backdoor Detection}\label{sec:evaldefense}

Our threat model (Section~\ref{sec:threat-model}) assumes the model tester uses state-of-the-art detection methods on submitted models to determine that they are not Trojaned before they are distributed through a model repository. The model deployer then downloads the tested model and compresses it for deployment, but does not perform their own backdoor detection tests on the compressed model. In this section, we evaluate the stealthiness of the injected backdoors when detection methods are used on the backdoored model in its uncompressed form. We emphasize that, we do not require the backdoors in the compressed model are stealthy, only that they are effective when compressed and stealthy for the uncompressed model. As we report in Section~\ref{sec:defense-method}, if the compressed models were tested as they will be deployed, the backdoor will be reliably detected. Our results support the potential for an adversary to hide a backdoor as a compression artifact. We test stealthiness of the artifact backdoor in the uncompressed model using two representative state-of-the-art backdoor detection methods: Meta Neural Trojan Detection (MNTD)~\cite{xu2019detecting} and Neural Cleanse~\cite{wang2019neural}.

\subsection{Meta Neural Trojan Detection}\label{sec:mntd-results}

Meta Neural Trojan Detection (MNTD)~\cite{xu2019detecting} uses meta classifiers to detect backdoors. It first trains many pairs of clean and backdoored models (generated by known backdoor attacks)
as the shadow models, and then uses these shadow models as the training set to build meta classifiers that can distinguish between clean and backdoored models. The trained meta classifier takes a model as input and outputs a score of maliciousness for that model, we name it as MNTD score, and smaller scores are expected for clean models and higher scores are expected for backdoored models. 

\shortsection{Experimental Setup}
Following the setup in the MNTD paper~\cite{xu2019detecting}, for each network architecture and dataset considered in Section \ref{sec:experiments}, we train 4,096 shadow models without considering model compression. Of these, 2,048 are clean models and the remainder are backdoored models with different trigger patterns and target classes. 
We then train five meta classifiers on the shadow models using different random seeds and report results averaged over the five meta classifiers. 
Following the original paper, we use the AUC that distinguishes the MNTD scores (generated by the meta classifier) of the clean models from those of the backdoored models to represent the detectability of our attack. A good attack strategy should produce lower AUC scores, ideally around or below 0.5, which means the backdoored (uncompressed) model is indistinguishable from the clean model. When computing the AUC using the trained meta-classifier, for each network architecture and dataset, we use 5 backdoored models trained with different target classes as instances in the positive class, and 10 clean models as instances in the negative class. We note that these ten clean models are not used in the model training involved in our attack process (described in Section~\ref{sec:attack-design}). In addition to AUC, we also show the True Positive Rate (TPR) in detecting the artifact backdoored models for a detector configuration set to bound the False Positive Rate (FPR) at 0.1 (that is, the highest detection threshold is used with the constraint that no more than 10\% of clean models are flagged as being backdoored) to give an intuitive understanding of the likelihood the backdoored model would be detected in a reasonable deployment.

\begin{table*}[htb!]
	\centering
	\footnotesize
	\begin{tabular}{cc|c|cc|cc|cc}
		\toprule
		\multirow{2}[4]{*}{Dataset} & \multirow{2}[4]{*}{Model} & \multicolumn{1}{c|}{\multirow{2}[4]{6em}{\centering Baseline AUC}} & \multicolumn{2}{p{13em}|}{\centering Known Pruning Rate} & \multicolumn{2}{p{13em}|}{\centering Unknown Pruning Rate} & \multicolumn{2}{c}{Quantization} \\
		&       &       & \scriptsize{Standard Attack} & \scriptsize{Distilled Attack} & \scriptsize{Standard Attack} & \scriptsize{Distilled Attack} & \scriptsize{Standard Attack} & \scriptsize{Distilled Attack} \\
		\midrule
		\multirow{3}[2]{*}{CIFAR-10} & VGG-16   & 0.6 $\pm$ 0.14 & 0.88 $\pm$ 0.12 & 0.62 $\pm$ 0.17 & 0.76 $\pm$ 0.21 & 0.60 $\pm$ 0.25 & 0.68 $\pm$ 0.28 & 0.49 $\pm$ 0.17 \\
		& ResNet-18 & 0.96 $\pm$ 0.04 & 0.81 $\pm$ 0.07 & 0.52 $\pm$ 0.22 & 0.92 $\pm$ 0.08 & 0.65 $\pm$ 0.18 & 0.74 $\pm$ 0.15 & 0.77 $\pm$ 0.05 \\
		& MobileNet & 0.99 $\pm$ 0.01 & 0.83 $\pm$ 0.15 & 0.62 $\pm$ 0.17 & 0.90 $\pm$ 0.12 & 0.84 $\pm$ 0.18 & 0.82 $\pm$ 0.20 & 0.87 $\pm$ 0.11 \\
		\midrule
		\multirow{3}[2]{*}{GTSRB} & VGG-16   & 0.83 $\pm$ 0.07 & 0.05 $\pm$ 0.03 & 0.72 $\pm$ 0.09 & 0.11 $\pm$ 0.10 & 0.46 $\pm$ 0.09 & 0.05 $\pm$ 0.02 & 0.70 $\pm$ 0.03 \\
		& ResNet-18 & 0.93 $\pm$ 0.03 & 0.01 $\pm$ 0.02 & 0.62 $\pm$ 0.10 & 0.04 $\pm$ 0.06 & 0.43 $\pm$ 0.12 & 0.19 $\pm$ 0.17 & 0.59 $\pm$ 0.13 \\
		& MobileNet & 0.81 $\pm$ 0.12 & 0.31 $\pm$ 0.20 & 0.58 $\pm$ 0.16 & 0.36 $\pm$ 0.18 & 0.44 $\pm$ 0.18 & 0.31 $\pm$ 0.21 & 0.62 $\pm$ 0.13 \\
		\bottomrule
	\end{tabular}%
	\caption{Detection AUC of MNTD.  \small The AUC score near or lower than 0.5 means the attack is undetectable.
		 Note that, the AUC score less than 0.5 means the detection is giving a completely wrong detection result (worse than random guessing). 
		 }	\label{tab:mntd-auc}%
\end{table*}%

\begin{table*}[htb!]
	\centering
	\footnotesize
	\begin{tabular}{cc|c|cc|cc|cc}
		\toprule
		\multirow{2}[4]{*}{Dataset} & \multirow{2}[4]{*}{Model} & \multicolumn{1}{c|}{\multirow{2}[4]{6em}{\centering Baseline TPR}} & \multicolumn{2}{p{13em}|}{\centering Known Pruning Rate} & \multicolumn{2}{p{13em}|}{\centering Unknown Pruning Rate} & \multicolumn{2}{c}{Quantization} \\
		&       &       & \scriptsize{Standard Attack} & \scriptsize{Distilled Attack} & \scriptsize{Standard Attack} & \scriptsize{Distilled Attack} & \scriptsize{Standard Attack} & \scriptsize{Distilled Attack} \\
		\midrule
		\multirow{3}[2]{*}{CIFAR-10} & VGG-16   & 0.36 $\pm$ 0.22 & 0.68 $\pm$ 0.32 & 0.40 $\pm$ 0.28 & 0.56 $\pm$ 0.32 & 0.40 $\pm$ 0.25 & 0.56 $\pm$ 0.37 & 0.32 $\pm$ 0.20 \\
		& ResNet-18 & 0.86 $\pm$ 0.19 & 0.48 $\pm$ 0.27 & 0.12 $\pm$ 0.16 & 0.76 $\pm$ 0.23 & 0.36 $\pm$ 0.23 & 0.48 $\pm$ 0.27 & 0.44 $\pm$ 0.08 \\
		& MobileNet & 0.98 $\pm$ 0.04 & 0.68 $\pm$ 0.32 & 0.32 $\pm$ 0.20 & 0.80 $\pm$ 0.25 & 0.68 $\pm$ 0.32 & 0.64 $\pm$ 0.32 & 0.68 $\pm$ 0.27 \\
		\midrule
		\multirow{3}[2]{*}{GTSRB} & VGG-16   & 0.64 $\pm$ 0.14 & 0.00 $\pm$ 0.00 & 0.48 $\pm$ 0.10 & 0.04 $\pm$ 0.08 & 0.20 $\pm$ 0.13 & 0.00 $\pm$ 0.00 & 0.44 $\pm$ 0.15 \\
		& ResNet-18 & 0.78 $\pm$ 0.07 & 0.00 $\pm$ 0.00 & 0.24 $\pm$ 0.15 & 0.00 $\pm$ 0.00 & 0.08 $\pm$ 0.10 & 0.08 $\pm$ 0.10 & 0.24 $\pm$ 0.23 \\
		& MobileNet & 0.68 $\pm$ 0.13 & 0.08 $\pm$ 0.10 & 0.36 $\pm$ 0.15 & 0.20 $\pm$ 0.13 & 0.20 $\pm$ 0.31 & 0.20 $\pm$ 0.22 & 0.48 $\pm$ 0.20 \\
		\bottomrule
	\end{tabular}%
	\caption{Detection TPRs of MNTD when FPRs are 0.1 (same experiments as Table~\ref{tab:mntd-auc}).}
	\label{tab:mntd-auc-tpr}%
\end{table*}%

\shortsection{Results on Regular Backdoor Attacks} We first confirm the effectiveness of the defense against regular backdoors injected without considering compression. For each network architecture and dataset, we train ten backdoored models (one for each target class) without considering model compression (following the training data poisoning method proposed by Chen et al.~\cite{chen2017targeted} and Gu et al.~\cite{gu2017badnets}). The ``Baseline AUC'' column in Table~\ref{tab:mntd-auc} shows the detection results for MNTD on the regular backdoored models, giving the AUC for distinguishing the backdoored models from the clean models.
From the table, we observe that MNTD is quite effective at detecting backdoors---the reported AUC scores are over 0.95 for the ResNet-18 and MobileNet CIFAR-10 models, and above 0.8 for all of the GTSRB models. The worst performance is for VGG (AUC = $0.6$) and it happens because the corresponding meta-classifier does not train properly in this setting.\footnote{We tried a few sets of hyperparameters, but all failed to make the meta-classifier work well in detecting the backdoors. We were not able to test more sets of hyperparameters because the meta-classifier training is very slow.} 
 
\shortsection{Results on Our Attacks}
Table~\ref{tab:mntd-auc} summarizes the detection performance of MNTD on the artifact backdoored models trained using the attacks described in Section~\ref{sec:quantization_experiments} and Section~\ref{sec:pruning_evaluation}. Overall, our attacks are significantly less detectable than the regular backdoor attacks. Our attacks are less detectable (e.g., lower AUC) than the baseline backdoor attacks on all the settings except the attack trained on VGG-16 for CIFAR-10, where the baseline detection is quite unsuccessful (AUC is 0.6), so in this case neither the baseline attack nor the artifact attacks are detected. 

The stealthiness of the attack can be better understood from Table~\ref{tab:mntd-auc-tpr}, which shows the detection rate for backdoored models when the threshold for the defense is set to a false positive rate of 10\%.  Of the six dataset-models and six attack methods, for 36 total configurations, the detection rate is below 10\% (that is, the backdoored models are less likely to be classified as malicious than the clean models are) for 8 of the 18 GTSRB settings. For the distilled attacks, the detection rate is below 50\% for all settings except for unknown pruning rate and quantization attacks on CIFAR-10 MobileNet. 




\shortersection{Settings where $AUC \ll 0.5$} For all the standard attacks on GTSRB, the AUC scores are below 0.5 and often close to 0.0, which means backdoored models resulted from our attacks have lower MNTD scores than those of clean models.
(see Figure~\ref{fig:roc-curves} in the Appendix for the actual AUC curves). 
In a setting where the defender knows the distribution of the models (half clean models and half artifact backdoored models), the classifier could be used as a nearly perfect detector of the artifact backdoored models by just flipping the output!  Of course, the actual defense is not in this setting. We hypothesized that the failure of MNTD in these settings is because the corresponding meta-classifiers are fooled by the effects of training the model for compression.
We validate this hypothesis by testing the same meta-classifier, but on normal clean models and clean models trained with compression. The clean model with compression is trained by removing the trigger pattern part in the loss function of the standard attack. 
For each setting, we train 5 clean models with compresssion and use the previous 10 normal clean models without compression to see if MNTD can distinguish them.
Table~\ref{tab:auc-clean-models} in the Appendix summarizes the results. 
For the CIFAR-10 models, the AUC scores are around 0.5, indicating that the meta-classifier is not distinguishing the models trained for compression. For GTSRB, however, the AUC scores are much smaller than 0.5 and show similar patterns to those of standard attack in Table~\ref{tab:mntd-auc}.  This supports our hypothesis that on GTSRB models the meta classifiers of MNTD just output low MNTD scores for models trained for compression, and including an artifact backdoor in this training still preserves the low scores which makes the artifact backdoored models less likely to be flagged than normal clean models. 

\shortersection{Distilled Attacks} 
The distilled attack is designed to be stealthier by producing artifact backdoored models that are similar to clean models. 
Experimental results confirm this for all settings of pruning attacks on CIFAR-10, but on GTSRB, we find the standard attacks are less detectable to MNTD. 

For CIFAR-10, all 6 standard pruning attack settings have AUC scores greater than 0.7 while for the distilled attacks only one setting has AUC higher than 0.7.  
The average detection rate (when the false positive rate is 10\%) is decreased from 66\% of standard attack to 38\% of distilled attack.
For the model quantization results, there is no clear winner and the distilled attack shows similar stealthiness to MNTD as the standard attack, and both attacks are relatively detectable--- the detection rate is higher than 50\% for half settings when the false positive rate is 10\%.
Decreasing the detection rate in model quantization is left as the future work. We also note that, although the detection rates of quantization settings are relatively high, they are still much smaller than the average  detection rate of 73\% of the baseline attacks. 

For GTSRB, all of our attacks are stealthy to MNTD, but the standard attacks are less detected than the distilled attacks. A possible reason for standard attack outperforming the distilled attack is, the standard attack has no constraint of pushing the model to behave similarly to the clean model, so it is more influenced by model compression during training (discussed in settings of AUC $\ll$ 0.5) while distilled attack forces the generated backdoored models to have similar properties as the clean models and pushes AUC scores near 0.5. However, having AUC near 0.5 is already very successful, indicating that the detection strategy cannot distinguish our backdoored models from the clean models.

\begin{table*}[tb]
	\centering
	\begin{tabular}{cc|c|cc|cc|cc}
		\toprule
		\multirow{2}[4]{*}{Dataset} & \multirow{2}[4]{*}{Model} & \multicolumn{1}{c|}{\multirow{2}[4]{6em}{\centering Baseline AUC}} & \multicolumn{2}{p{13em}|}{\centering Known Pruning Rate} & \multicolumn{2}{p{13em}|}{\centering Unknown Pruning Rate} & \multicolumn{2}{c}{Quantization} \\
		&       &       & \scriptsize{Standard Attack} & \scriptsize{Distilled Attack} & \scriptsize{Standard Attack} & \scriptsize{Distilled Attack} & \scriptsize{Standard Attack} & \scriptsize{Distilled Attack} \\
		\midrule
		\multirow{3}[2]{*}{CIFAR-10} & VGG-16   & 0.9 (0.9)  & 0.38 (0.06)  & 0.54 (0.12)  & 0.76 (0.38)  & 0.32 (0.08)  & 0.42 (0.26)  & 0.48 (0.36) \\
		& ResNet-18 & 1 (1)    & 0.56 (0.24)  & 0.42 (0.08)  & 0.7 (0.28)   & 0.66 (0.28)  & 0.72 (0.24)  & 0.6 (0.1) \\
		& MobileNet & 0.8 (0.73)  & 0.48 (0.14) & 0.52 (0.16)  & 0.6 (0)  & 0.2 (0)  & 0.32 (0.06) & 0.52 (0.14) \\
		\midrule
		\multirow{3}[2]{*}{GTSRB} & VGG-16   & 0.98 (0.98)  & 0.72 (0.18)  & 0.72 (0.18)  & 0.58 (0.18)  & 0.74 (0.18) & 0.67 (0) & 0.58 (0)\\
		& ResNet-18 & 1  (1)   & 0.9 (0.2)  & 0.66 (0)  & 1 (0.2)    & 0.6 (0.4)   & 0.54 (0) & 0.2 (0) \\
		& MobileNet & 0.84 (0.84)  & 0.08 (0.04) & 0.18 (0)  & 0.18 (0)  & 0 (0)     & 0.38 (0)  & 0.18 (0) \\
		\bottomrule
	\end{tabular}%
	\caption{Detection AUC of Neural Cleanse. \small The AUC values outside the parenthesis show the results of normal detection settings where the model tester only cares about the maximum anomaly index of a given model. The AUC values inside the parenthesis are evaluated differently---the anomaly indexes of the target classes of backdoored models are used as the positive class. 
	We show the anomaly indexes of some backdoored models as examples in Appendix Figure~\ref{fig:example-heat-maps}. We also include the TPR when the FPR is 0.1 in Table~\ref{tab:neural-cleanse-detection-results-tpr} in the Appendix.}
	
	\label{tab:neural-cleanse-detection-results}%
\end{table*}%

\begin{table*}[htb!]
	\centering
	\footnotesize
	\begin{tabular}{c|cc|cc|cc}
		\toprule
		\multirow{2}[4]{*}{CIFAR-10 Model} & \multicolumn{2}{c|}{\centering Known Pruning Rate} & \multicolumn{2}{c|}{Unknown Pruning Rate} & \multicolumn{1}{c}{Quantization} \\
		& MNTD & Neural Cleanse & MNTD  & Neural Cleanse & MNTD \\ 
		\midrule
		VGG   & 0.89 $\pm$ 0.14 & 0.78 (0.78) & 0.68 $\pm$ 0.21 & 0.18 (0.02) & 0.89 $\pm$ 0.12 \\ 
		ResNet-18 & 1.00 $\pm$ 0.00 & 1.00 (1.00) & 1.00 $\pm$ 0.00 & 0.98 (0.98) & 0.99 $\pm$ 0.02 \\ 
		MobileNet & 1.00 $\pm$ 0.00 & 0.84 (0.84) & 1.00 $\pm$ 0.00 & 0.80 (0.80) & 1.00 $\pm$ 0.00 \\ 
		\bottomrule
	\end{tabular}%
	\caption{Detection AUC after Compression. \small Artifact backdoor models are from distilled attacks on CIFAR-10. 
	For known pruning rate setting, we compress models with pruning rate of 0.3. For unknown pruning rate setting, we compress models with pruning rate of 0.4. We run detection tools taking compressed model as inputs. or model pruning, since the pruned model is in the same format as a normally trained model, both detection work without modification. For model quantization, due to the zero-gradient issue, we only include MNTD (it may be possible to use the STE adopted in QAT (Section~\ref{sec:8-bit-quantization-brief}) to tackle this problem, but we have not tested this).}
	\label{tab:detecting-compressed-models}%
\end{table*}%

\subsection{Neural Cleanse}\label{sec:neural-cleanse-evaluation}


Recall (from Section~\ref{sec:background-backdoors}) that Neural Cleanse~\cite{wang2019neural} detects backdoored models by reconstructing trigger patterns for each output class. If there is a small trigger pattern for an output class, this is an evidence of a backdoored model and the output class with smallest reconstructed trigger size is likely to be the backdoor target class. Neural Cleanse uses gradient descent to reconstruct the trigger patterns. To avoid ending in bad local optima, we repeat the whole process of trigger reconstruction three times with different initializations, and keep the smallest trigger pattern found from the three trials. 

After computing the reversed triggers for all output classes, Neural Cleanse generates an anomaly index (a normalized median absolute deviation value of the $\ell_1$-norm of triggers) for each output class. The authors use an anomaly index of $2$ as the threshold for declaring a backdoor, and the class with smallest $\ell_1$-norm and an anomaly index over $2$ is reported as the target class~\cite{wang2019neural}. We find that using the anomaly index threshold of 2 sometimes leads to very high false positive rates. For example, 6 of 10 normal clean MobileNet models trained on GTSRB are classified as backdoored (Appendix Figure~\ref{fig:neural_cleanse_normal_clean_model_mobilenet_gtsrb}). Since setting the right threshold for the anomaly index is somewhat arbitrary, similar to MNTD, we report the AUC on distinguishing the maximum anomaly indexes of clean models from those of backdoored models to avoid the thresholding issues. A stealthier attack should have a lower AUC value, and a score close to 0.5 indicates the detector is not doing better than random guessing in distinguishing clean models from the artifact backdoored models.

\shortsection{Results} 
Table~\ref{tab:neural-cleanse-detection-results} summarizes the detection performance of Neural Cleanse. The ``Baseline AUC'' column shows the detection AUC for Neural Cleanse on the regular backdoor models. The results are very competitive---the AUC values are all above 0.8 for all settings and achieves 1.0 (perfect detection) for ResNet-18 on both datasets, which means that Neural Cleanse is quite effective at detecting regular backdoors. The AUC scores for our artifact backdoor attacks are lower than those of the baseline attacks in all attack settings.
While all baseline AUCs are greater than 0.8, the AUC for the artifact backdoors is below 0.8 for all but one (standard attack on ResNet-18 and GTSRB) of the 36 attack settings. To better understand the AUC values, 
consider a model tester constrained by a 10\% maximum false positive rate (details in Table~\ref{tab:neural-cleanse-detection-results-tpr} in the Appendix)---for 9 of the 36 attack settings, the detection rate would be 0\%.
For model pruning, the average detection rate for distilled attack models (at false positive rate of 10\%) is 28\% (compared to 35\% for the standard attack); for quantization, the detection rate is 17\% (compared to 30\% for the standard attack).  



\shortersection{Identifying the Target Class} Although Neural Cleanse produces relatively high AUC scores on small fraction of the attack settings (7 of 36 attack settings have AUC scores higher than 0.7), the detection strategy fails to identify the actual target class used for backdoor attacks, which is the purpose of the original paper. We report in Table~\ref{tab:neural-cleanse-detection-results} the AUC for both detecting any backdoor and, in parentheses, for identifying the correct target class. The original Neural Cleanse paper consider detection successful only when the actual target class is identified. The AUC scores for correct target class identification are significantly lower than those outside the parenthesizes, which indicates the actual target class of the backdoored model is not the one with maximum anomaly index. The extremely small AUC values outside the parenthesizes (e.g., 0 for MobileNet on GTSRB) also suggest the anomaly indices of the artifact backdoored models are even smaller than those of the normal clean models and performs worse than random guessing. We still hypothesize this is due to the failure of Neural Cleanse in properly handling clean models trained with compression. 
We ran similar experiments to those for the MNTD case on Neural Cleanse and also observe extremely small AUC scores when distinguishing clean models with compression and normal clean models without compression. 

	\section{Defending against Artifact Attacks} \label{sec:defense-method}


A simple defense against compression artifact attacks is to ensure that model testing is done on the model as it will be deployed, not on a pre-compressed model. Table~\ref{tab:detecting-compressed-models} summarizes the results of our direct defense experiments, showing that both MNTD and Neural Cleanse are able to detect the backdoor when testing the compressed model. For MNTD, for five of the nine settings the detection is perfect, and for all but one setting (VGG with unknown pruning rate) it is at least 0.89. For Neural Cleanse, five of the six settings have AUC scores greater than 0.78.





	\section{Conclusion}
We introduce a new kind of stealthy backdoor attack on deep learning classifiers that hides a backdoor as a compression artifact. We design and demonstrate the effectiveness of our attack methods for the two most common model compression techniques---model pruning and model quantization, and also against the state-of-the-art detection methods. Our attacks have the flavor of ``time-of-check, time-of-use" vulnerabilities, a problem well understood for many decades~\cite{abbott1976security, bishop1995race}, and reinforce a lesson that has been learned in many security domains: any gap between the artifact which is tested for security and the instantiation as used presents an opportunity for attackers to exploit.


	\section*{Acknowledgements}

This work was partially funded by awards from the NSFC-61872180, the National Science Foundation (NSF) SaTC program (Center for Trustworthy Machine Learning, \#1804603), Jiangsu "Shuang-Chuang" Program, and Jiangsu "Six-Talent-Peaks" Program.

	\bibliographystyle{IEEEtran}
	\bibliography{invisible_backdoor}

\begin{thebibliography}{10}
\providecommand{\url}[1]{#1}
\csname url@samestyle\endcsname
\providecommand{\newblock}{\relax}
\providecommand{\bibinfo}[2]{#2}
\providecommand{\BIBentrySTDinterwordspacing}{\spaceskip=0pt\relax}
\providecommand{\BIBentryALTinterwordstretchfactor}{4}
\providecommand{\BIBentryALTinterwordspacing}{\spaceskip=\fontdimen2\font plus
\BIBentryALTinterwordstretchfactor\fontdimen3\font minus
  \fontdimen4\font\relax}
\providecommand{\BIBforeignlanguage}[2]{{%
\expandafter\ifx\csname l@#1\endcsname\relax
\typeout{** WARNING: IEEEtran.bst: No hyphenation pattern has been}%
\typeout{** loaded for the language `#1'. Using the pattern for}%
\typeout{** the default language instead.}%
\else
\language=\csname l@#1\endcsname
\fi
#2}}
\providecommand{\BIBdecl}{\relax}
\BIBdecl

\bibitem{simonyan2014very}
K.~Simonyan and A.~Zisserman, ``Very deep convolutional networks for
  large-scale image recognition,'' in \emph{ICLR}, 2015.

\bibitem{he2016deep}
K.~He, X.~Zhang, S.~Ren, and J.~Sun, ``Deep residual learning for image
  recognition,'' in \emph{CVPR}, 2016.

\bibitem{huang2017densely}
G.~Huang, Z.~Liu, L.~van~der Maaten, and K.~Q. Weinberger, ``Densely connected
  convolutional networks,'' in \emph{CVPR}, 2017.

\bibitem{chandrasekhar2017compression}
V.~Chandrasekhar, J.~Lin, Q.~Liao, O.~Morere, A.~Veillard, L.~Duan, and
  T.~Poggio, ``Compression of deep neural networks for image instance
  retrieval,'' in \emph{DCC}, 2017.

\bibitem{brown2020language}
T.~B. Brown, B.~Mann, N.~Ryder, M.~Subbiah, J.~Kaplan, P.~Dhariwal,
  A.~Neelakantan, P.~Shyam, G.~Sastry, A.~Askell \emph{et~al.}, ``Language
  models are few-shot learners,'' \emph{arXiv:2005.14165}, 2020.

\bibitem{migacz20178}
S.~Migacz, ``8-bit inference with {TensorRT},'' in \emph{GTC}, 2017.

\bibitem{jacob2018quantization}
B.~Jacob, S.~Kligys, B.~Chen, M.~Zhu, M.~Tang, A.~Howard, H.~Adam, and
  D.~Kalenichenko, ``Quantization and training of neural networks for efficient
  integer-arithmetic-only inference,'' in \emph{CVPR}, 2018.

\bibitem{dong2017learning}
X.~Dong, S.~Chen, and S.~Pan, ``Learning to prune deep neural networks via
  layer-wise optimal brain surgeon,'' in \emph{NeurIPS}, 2017.

\bibitem{srinivas2015data}
S.~Srinivas and R.~V. Babu, ``Data-free parameter pruning for deep neural
  networks,'' \emph{arXiv:1507.06149}, 2015.

\bibitem{han2015learning}
S.~Han, J.~Pool, J.~Tran, and W.~Dally, ``Learning both weights and connections
  for efficient neural network,'' in \emph{NeurIPS}, 2015.

\bibitem{guo2016dynamic}
Y.~Guo, A.~Yao, and Y.~Chen, ``Dynamic network surgery for efficient {DNNs},''
  in \emph{NeurIPS}, 2016.

\bibitem{lin2018synaptic}
C.~Lin, Z.~Zhong, W.~Wei, and J.~Yan, ``Synaptic strength for convolutional
  neural network,'' in \emph{NeurIPS}, 2018.

\bibitem{wen2016learning}
W.~Wen, C.~Wu, Y.~Wang, Y.~Chen, and H.~Li, ``Learning structured sparsity in
  deep neural networks,'' in \emph{NeurIPS}, 2016.

\bibitem{figurnov2016perforatedcnns}
M.~Figurnov, A.~Ibraimova, D.~P. Vetrov, and P.~Kohli, ``Perforatedcnns:
  Acceleration through elimination of redundant convolutions,'' in
  \emph{NeurIPS}, 2016.

\bibitem{li2016pruning}
H.~Li, A.~Kadav, I.~Durdanovic, H.~Samet, and H.~P. Graf, ``Pruning filters for
  efficient convnets,'' \emph{arXiv:1608.08710}, 2016.

\bibitem{yang2017designing}
T.-J. Yang, Y.-H. Chen, and V.~Sze, ``Designing energy-efficient convolutional
  neural networks using energy-aware pruning,'' in \emph{CVPR}, 2017.

\bibitem{liu2020autocompress}
N.~Liu, X.~Ma, Z.~Xu, Y.~Wang, J.~Tang, and J.~Ye, ``Autocompress: An automatic
  dnn structured pruning framework for ultra-high compression rates,'' in
  \emph{AAAI}, 2020.

\bibitem{krishnamoorthi2018quantizing}
R.~Krishnamoorthi, ``Quantizing deep convolutional networks for efficient
  inference: A whitepaper,'' \emph{arXiv:1806.08342}, 2018.

\bibitem{paszke2019pytorch}
A.~Paszke, S.~Gross, F.~Massa, A.~Lerer, J.~Bradbury, G.~Chanan, T.~Killeen,
  Z.~Lin, N.~Gimelshein, L.~Antiga \emph{et~al.}, ``Pytorch: An imperative
  style, high-performance deep learning library,'' in \emph{NeurIPS}, 2019.

\bibitem{abadi2016tensorflow}
M.~Abadi, P.~Barham, J.~Chen, Z.~Chen, A.~Davis, J.~Dean, M.~Devin,
  S.~Ghemawat, G.~Irving, M.~Isard \emph{et~al.}, ``Tensorflow: A system for
  large-scale machine learning,'' in \emph{OSDI}, 2016.

\bibitem{vanholder2016efficient}
H.~Vanholder, ``Efficient inference with {TensorRT},'' {GTC-EU} Presentation,
  2017.

\bibitem{coreml}
{Apple, Inc.}, ``Core {ML},''
  \url{https://developer.apple.com/documentation/coreml}.

\bibitem{liu2017trojaning}
Y.~Liu, S.~Ma, Y.~Aafer, W.-C. Lee, J.~Zhai, W.~Wang, and X.~Zhang, ``Trojaning
  attack on neural networks,'' in \emph{NDSS}, 2018.

\bibitem{gu2017badnets}
T.~Gu, B.~Dolan-Gavitt, and S.~Garg, ``{BadNets}: Identifying vulnerabilities
  in the machine learning model supply chain,'' \emph{arXiv:1708.06733}, 2017.

\bibitem{modelzoo}
J.~Y. Koh, ``{ModelZoo},'' \url{https://modelzoo.co}, 2020.

\bibitem{devlin2018bert}
J.~Devlin, M.-W. Chang, K.~Lee, and K.~Toutanova, ``{BERT}: Pre-training of
  deep bidirectional transformers for language understanding,''
  \emph{arXiv:1810.04805}, 2018.

\bibitem{feng2020codebert}
Z.~Feng, D.~Guo, D.~Tang, N.~Duan, X.~Feng, M.~Gong, L.~Shou, B.~Qin, T.~Liu,
  D.~Jiang \emph{et~al.}, ``Codebert: A pre-trained model for programming and
  natural languages,'' \emph{arXiv:2002.08155}, 2020.

\bibitem{kolesnikov2019big}
A.~Kolesnikov, L.~Beyer, X.~Zhai, J.~Puigcerver, J.~Yung, S.~Gelly, and
  N.~Houlsby, ``Big transfer ({BiT}): General visual representation learning,''
  \emph{arXiv:1912.11370}, 2019.

\bibitem{xie2020self}
Q.~Xie, M.-T. Luong, E.~Hovy, and Q.~V. Le, ``Self-training with noisy student
  improves imagenet classification,'' in \emph{CVPR}, 2020.

\bibitem{touvron2020fixing}
H.~Touvron, A.~Vedaldi, M.~Douze, and H.~J{\'e}gou, ``Fixing the train-test
  resolution discrepancy: {FixEfficientNet},'' \emph{arXiv:2003.08237}, 2020.

\bibitem{msr-secureai}
A.~Marshall, R.~Rojas, J.~Stokes, and D.~Brinkman, ``Securing the future of
  artificial intelligence and machine learning at {M}icrosoft,''
  \url{https://docs.microsoft.com/en-us/security/engineering/securing-artificial-intelligence-machine-learning},
  2020.

\bibitem{wang2019neural}
B.~Wang, Y.~Yao, S.~Shan, H.~Li, B.~Viswanath, H.~Zheng, and B.~Y. Zhao,
  ``Neural cleanse: Identifying and mitigating backdoor attacks in neural
  networks,'' in \emph{IEEE {S\&P}}, 2019.

\bibitem{gao2019strip}
Y.~Gao, C.~Xu, D.~Wang, S.~Chen, D.~C. Ranasinghe, and S.~Nepal, ``Strip: A
  defence against {T}rojan attacks on deep neural networks,'' in \emph{ACSAC},
  2019.

\bibitem{liu2019abs}
Y.~Liu, W.-C. Lee, G.~Tao, S.~Ma, Y.~Aafer, and X.~Zhang, ``{ABS}: Scanning
  neural networks for back-doors by artificial brain stimulation,'' in
  \emph{CCS}, 2019.

\bibitem{chen2019deepinspect}
H.~Chen, C.~Fu, J.~Zhao, and F.~Koushanfar, ``{DeepInspect}: A black-box trojan
  detection and mitigation framework for deep neural networks.'' in
  \emph{IJCAI}, 2019.

\bibitem{gong2014compressing}
Y.~Gong, L.~Liu, M.~Yang, and L.~Bourdev, ``Compressing deep convolutional
  networks using vector quantization,'' \emph{arXiv:1412.6115}, 2014.

\bibitem{wu2016quantized}
J.~Wu, C.~Leng, Y.~Wang, Q.~Hu, and J.~Cheng, ``Quantized convolutional neural
  networks for mobile devices,'' in \emph{CVPR}, 2016.

\bibitem{choi2016towards}
Y.~Choi, M.~El-Khamy, and J.~Lee, ``Towards the limit of network
  quantization,'' \emph{arXiv:1612.01543}, 2016.

\bibitem{halfquantization}
{TensorFlow Team}, ``Tensorflow model optimization toolkit — float16
  quantization halves model size,''
  \url{https://blog.tensorflow.org/2019/08/tensorflow-model-optimization-toolkit_5.html}.

\bibitem{trtorch_post_training_quantization}
{NVidia, Inc.}, ``Post training quantization of {TRTorch},''
  \url{https://nvidia.github.io/TRTorch/tutorials/ptq.html}.

\bibitem{tensorflow_post_training_quantization}
{TensorFlow Team}, ``Post-training quantization of {TensorFlow},''
  \url{https://www.tensorflow.org/lite/performance/post_training_quantization}.

\bibitem{bengio2013estimating}
Y.~Bengio, N.~L{\'e}onard, and A.~Courville, ``Estimating or propagating
  gradients through stochastic neurons for conditional computation,''
  \emph{arXiv:1308.3432}, 2013.

\bibitem{alvarez2016learning}
J.~M. Alvarez and M.~Salzmann, ``Learning the number of neurons in deep
  networks,'' in \emph{NeurIPS}, 2016.

\bibitem{he2018amc}
Y.~He, J.~Lin, Z.~Liu, H.~Wang, L.-J. Li, and S.~Han, ``Amc: Automl for model
  compression and acceleration on mobile devices,'' in \emph{ECCV}, 2018.

\bibitem{chen2017targeted}
X.~Chen, C.~Liu, B.~Li, K.~Lu, and D.~Song, ``Targeted backdoor attacks on deep
  learning systems using data poisoning,'' \emph{arXiv:1712.05526}, 2017.

\bibitem{guo2019tabor}
W.~Guo, L.~Wang, X.~Xing, M.~Du, and D.~Song, ``Tabor: A highly accurate
  approach to inspecting and restoring trojan backdoors in ai systems,'' in
  \emph{ICDM}, 2019.

\bibitem{huang2019neuroninspect}
X.~Huang, M.~Alzantot, and M.~Srivastava, ``{NeuronInspect}: Detecting
  backdoors in neural networks via output explanations,'' in \emph{AAAI}, 2019.

\bibitem{xu2019detecting}
X.~Xu, Q.~Wang, H.~Li, N.~Borisov, C.~A. Gunter, and B.~Li, ``Detecting {AI}
  {T}rojans using meta neural analysis,'' in \emph{IEEE {S\&P}}, 2021.

\bibitem{yao2019latent}
Y.~Yao, H.~Li, H.~Zheng, and B.~Y. Zhao, ``Latent backdoor attacks on deep
  neural networks,'' in \emph{CCS}, 2019.

\bibitem{tang2020embarrassingly}
R.~Tang, M.~Du, N.~Liu, F.~Yang, and X.~Hu, ``An embarrassingly simple approach
  for {Trojan} attack in deep neural networks,'' in \emph{KDD}, 2020.

\bibitem{salem2020dynamic}
A.~Salem, R.~Wen, M.~Backes, S.~Ma, and Y.~Zhang, ``Dynamic backdoor attacks
  against machine learning models,'' \emph{arXiv:2003.03675}, 2020.

\bibitem{xiao2019seeing}
Q.~Xiao, Y.~Chen, C.~Shen, Y.~Chen, and K.~Li, ``Seeing is not believing:
  Camouflage attacks on image scaling algorithms,'' in \emph{USENIX Security},
  2019.

\bibitem{ye2019adversarial}
S.~Ye, K.~Xu, S.~Liu, H.~Cheng, J.-H. Lambrechts, H.~Zhang, A.~Zhou, K.~Ma,
  Y.~Wang, and X.~Lin, ``Adversarial robustness vs.\ model compression, or
  both?'' in \emph{ICCV}, 2019.

\bibitem{pytorchquantization}
{Torch Contributors}, ``{PyTorch} quantization,''
  \url{https://pytorch.org/docs/stable/quantization.html}.

\bibitem{mo2020darknetz}
F.~Mo, A.~S. Shamsabadi, K.~Katevas, S.~Demetriou, I.~Leontiadis, A.~Cavallaro,
  and H.~Haddadi, ``{DarkneTZ}: towards model privacy at the edge using trusted
  execution environments,'' in \emph{MobiSys}, 2020.

\bibitem{kim2020vessels}
K.~Kim, C.~H. Kim, J.~J. Rhee, X.~Yu, H.~Chen, D.~Tian, and B.~Lee, ``Vessels:
  efficient and scalable deep learning prediction on trusted processors,'' in
  \emph{ACM Symposium on Cloud Computing}, 2020.

\bibitem{hinton2015distilling}
G.~Hinton, O.~Vinyals, and J.~Dean, ``Distilling the knowledge in a neural
  network,'' \emph{arXiv:1503.02531}, 2015.

\bibitem{krizhevsky2014cifar}
A.~Krizhevsky, V.~Nair, and G.~Hinton, ``The {CIFAR} dataset.''

\bibitem{Houben-IJCNN-2013}
S.~Houben, J.~Stallkamp, J.~Salmen, M.~Schlipsing, and C.~Igel, ``Detection of
  traffic signs in real-world images: The {G}erman {T}raffic {S}ign {D}etection
  {B}enchmark,'' in \emph{IJCNN}, 2013.

\bibitem{howard2017mobilenets}
A.~G. Howard, M.~Zhu, B.~Chen, D.~Kalenichenko, W.~Wang, T.~Weyand,
  M.~Andreetto, and H.~Adam, ``{MobileNets}: Efficient convolutional neural
  networks for mobile vision applications,'' \emph{arXiv:1704.04861}, 2017.

\bibitem{tsrd}
L.~Huang, ``Chinese traffic sign database,''
  \url{https://www.nlpr.ia.ac.cn/pal/trafficdata/recognition.html}.

\bibitem{netzer2011reading}
Y.~Netzer, T.~Wang, A.~Coates, A.~Bissacco, B.~Wu, and A.~Y. Ng, ``Reading
  digits in natural images with unsupervised feature learning,'' in
  \emph{NeurIPS}, 2011.

\bibitem{abbott1976security}
R.~P. Abbott, J.~S. Chin, J.~E. Donnelley, W.~L. Konigsford, S.~Tokubo, and
  D.~A. Webb, ``Security analysis and enhancements of computer operating
  systems,'' National Bureau of Standards, Tech. Rep., 1976.

\bibitem{bishop1995race}
M.~Bishop, ``Race conditions, files, and security flaws; or the tortoise and
  the hare redux,'' UC Davis, Tech. Rep., 1995.

\bibitem{filterprungl2}
{Neural Network Intelligence}, ``Filter-level structured pruning based on the
  $\ell_2$-norm,''
  \url{https://nni.readthedocs.io/en/stable/Compression/Pruner.html#l1filter-pruner},
  2021.

\end{thebibliography}
	\clearpage
	\onecolumn
\appendix

\renewcommand\thefigure{\Alph{section}\arabic{figure}}    
\setcounter{figure}{0} 

\renewcommand\thetable{\Alph{section}\arabic{table}}    
\setcounter{table}{0}

\setcounter{equation}{0}
\renewcommand\theequation{A.\arabic{equation}}

\subsection{Model Quantization Details} \label{sec:quantization-details}
This section provides more details regarding the model quantization technique used in the paper. Eight-bit quantization impacts both training and inference. We covered training in  Section~\ref{sec:8-bit-quantization-brief}. This section covers how to perform the inference efficiently for 8-bit quantized models.

Matrix multiplication operations consume the most computing resources in deep learning model inference. Here we show how FP32 matrix multiplication $ A = B \times C$ is optimized under 8-bit precision:
\begin{align}
A &= B \times C \\
&= s_B (B_Q - z_B)  s_C (C_Q - z_C) \\
&= s_B s_C \left[ B_Q C_Q - C_Q z_B  - B_Q z_C + z_B z_C \right] \label{eq:outputfp32} \\   
s_A (A_Q - z_A)  &= s_B s_C \left[ B_Q C_Q - C_Q z_B  - B_Q z_C + z_B z_C \right] \\
A_Q &= \frac{s_B s_C }{s_A}\left[ B_Q C_Q - C_Q z_B  - B_Q z_C \right] + \frac{s_B s_C }{s_A} z_B z_C  + z_A  \label{eq:quant_expanded}
\end{align}
where $A_Q$, $B_Q$, and $C_Q$ are the quantized versions of $A$, $B$, and $C$, respectively. $s_A$, $s_B$, and $s_C$ are the quantization parameters ($s$ in~\eqref{eq:def_quant_affine}) for quantizing $A$, $B$, and $C$, respectively. And $z_A$, $z_B$, and $z_C$ are the corresponding quantization parameter $z$ in~\eqref{eq:def_quant_affine}.

The matrix multiplication can be accelerated by using \eqref{eq:outputfp32}. The calculations in the brackets are all performed with 8-bit integers using  8-bit or 16-bit integer instructions, which speeds up the computation. The multiplier $s_B$ and $s_C$ can be precalculated because all quantization parameters are obtained before model inference. Equation~\eqref{eq:quant_expanded} can further optimize the matrix multiplication by directly outputting the quantized multiplication result for use in the next matrix multiplication. Note that the $\frac{s_B s_C }{s_A}$ and  $\frac{s_B s_C }{s_A} z_B z_C  + z_A$ in this equation can also be precalculated to save the computation time. In addition, \eqref{eq:quant_expanded} can be adapted to support full integer mode to achieve even better performance~\cite{jacob2018quantization}.

\subsection{Effect of disabling bias of Conv2d layer of VGG-16} \label{sec:study_impact_of_change}

To be compatible with QAT, we needed to disable the bias option of the Conv2D layer of VGG-16. To study the impact of this modification on model accuracy, we use the original VGG-16 to train ten clean models for CIFAR-10 and GTSRB separately, and compare the averaged accuracies with those of the modified VGG-16 models. Table \ref{tab:enable_bias_option} shows the results---removing the bias has little impact on VGG-16 model performance. 

\begin{table}[!htb]
	\centering
	\footnotesize
	\begin{tabular}{ccc}
		\toprule
		Dataset & {\centering Model trained with bias} & {\centering Model trained without bias} \\
		\midrule
		CIFAR-10 & 92.8 $\pm$ 0.2 & 92.9 $\pm$ 0.2 \\
		GTSRB & 97.8 $\pm$ 0.4 & 97.7 $\pm$ 0.3 \\
		\bottomrule
	\end{tabular}%
	\caption{VGG-16 models trained with and without bias.}
	\label{tab:enable_bias_option}%
\end{table}%

\begin{table*}[tbh!]
	\centering
	\footnotesize

	\begin{tabular}{cc|cc|ccc|c}
		\toprule
		\multirow{3}[6]{*}{Dataset} & \multirow{3}[6]{*}{Model} & \multicolumn{5}{c}{Quantizing all the layers} & Quantizing last few layers \\
		&       & \multicolumn{2}{c}{Uncompressed Backdoored Model} & \multicolumn{3}{c}{Compressed Backdoored Model} & Compressed Backdoored Model  \\
		&       & {\centering Accuracy } & \multicolumn{1}{c}{\centering Triggered Accuracy} & {\centering Accuracy} & {\centering Triggered Accuracy} & \multicolumn{1}{c}{\centering Attack Success} & {Attack Success} \\
		\midrule
		\multirow{2}[2]{*}{CIFAR-10} & ResNet-18 & 92.1 $\pm$ 0.4 & 91.9 $\pm$ 0.3 & 91.9 $\pm$ 0.4 & 58.0 $\pm$ 38.7 & 42.3 $\pm$ 46.6 & 99.1 $\pm$ 1.3 \\
		& MobileNet & 92.0 $\pm$ 0.2 & 91.7 $\pm$ 0.2 & 91.7 $\pm$ 0.3 & 27.2 $\pm$ 31.9 & 79.0 $\pm$ 39.1 & 98.4 $\pm$ 2.6 \\
		\midrule
		\multirow{2}[2]{*}{GTSRB} & ResNet-18 & 97.5 $\pm$ 0.2 & 97.5 $\pm$ 0.2 & 97.2 $\pm$ 0.1 & 2.3 $\pm$ 1.6 & 99.9 $\pm$ 0.1 & 100.0 $\pm$ 0.0 \\
		& MobileNet & 98.0 $\pm$ 0.2 & 98.0 $\pm$ 0.2 & 97.9 $\pm$ 0.2 & 52.0 $\pm$ 42.0 & 48.8 $\pm$ 44.5 & 96.6 $\pm$ 6.5 \\
		\bottomrule
	\end{tabular}%
	
	\caption{Quantizing all layers compared to quantizing just the last few layers in model training.}
\label{tab:impact-number-of-layers-quantization}%
\end{table*}%

\subsection{Quantization Strategy in Model Training} \label{sec:quantization-strategy-in-model-training}

In this section, we show that, for model quantization,
although the model deployer will quantize all layers of a model to reduce the model size maximally, the attacker does not need to quantize all layers of the model when generating compressed models during training. In fact, we find that quantizing all the layers during training often lowers attack effectiveness. 
Table~\ref{tab:impact-number-of-layers-quantization} shows the results of uantize all the layers of ResNet-18 and MobileNet when generating compressed models during model training and all the training under the framework of standard attack (compare to the settings in Section~\ref{sec:known-calibration-dataset} where only some layers are quantized). Quantizing all the layers of a model results in models with good clean accuracies in both their uncompressed and compressed forms, but the attack success rates drop a lot compared to those of the training strategy which only quantizes the last few layers in the model training.
The attack success rates on three of the four settings drops by 19--56\%. So, quantization backdoors are most effective when 
the attacker only quantizes the last few layers of the model in the training, even though the model deployer will quantize all the layers to maximize compression.

\subsection{Results of Standard Attack}
\label{sec:standard-attack-results}
\begin{figure*}[!t]
	\centering
	\includegraphics[width=0.9\textwidth]{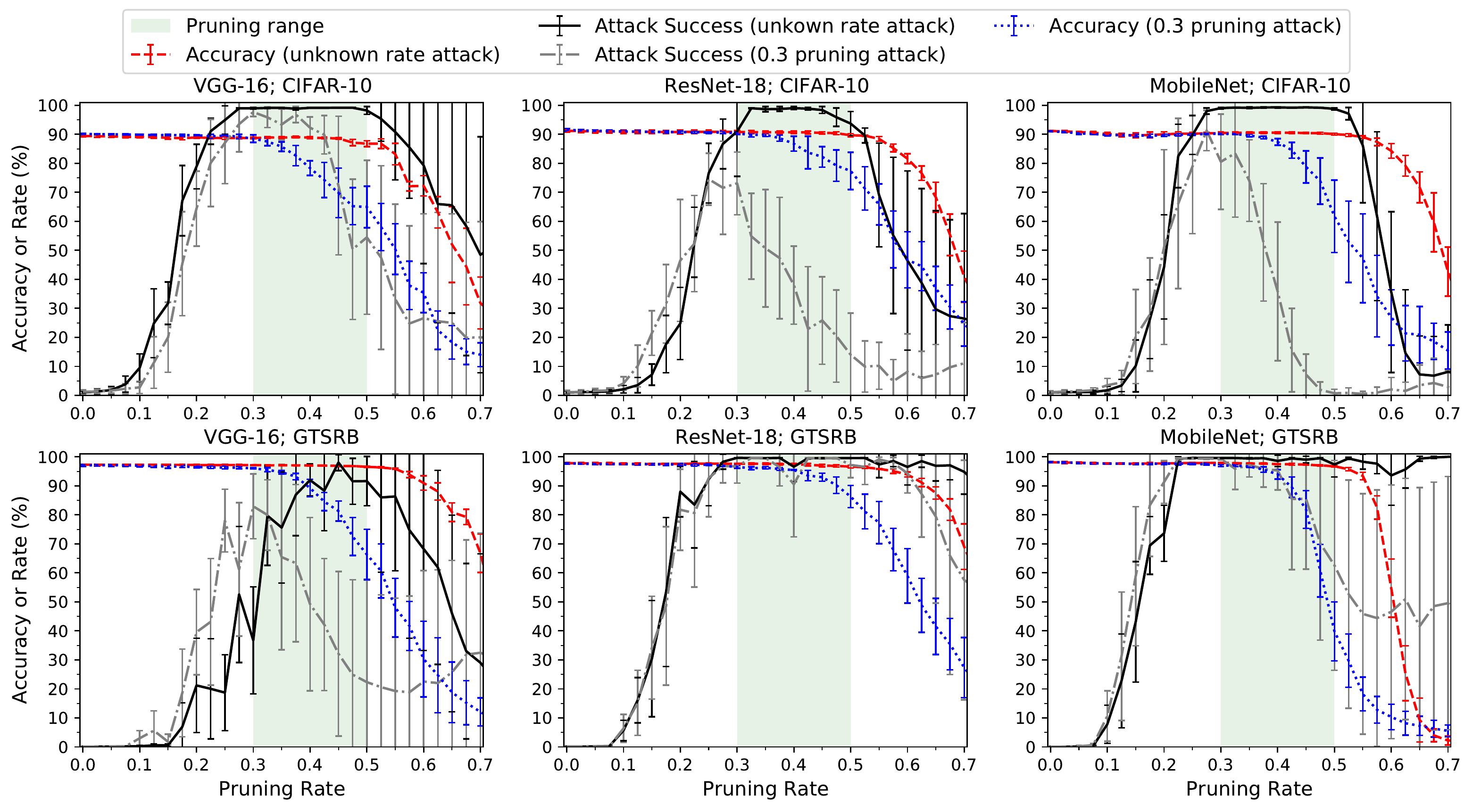}
	
	\caption{Effectiveness of standard pruning attacks across range of pruning rates. \small The uncompressed backdoor models are pruned over a range of victim pruning rates (from 0.0 (no pruning) to 0.7). The shaded green areas show the pruning range (0.3--0.5) targeted by the adversary in injecting the backdoor.
}
	\label{fig:standard_pruning_ranges_attack_effectiveness}
\end{figure*}

The results in the main body focus on the distilled attack because of its better stealthiness. Here, we show results of the standard attacks, which tend to be slightly more effective than the distilled attacks disregarding the detection risk. Table~\ref{tab:normal-calibration-dataset} shows the effectiveness of the standard quantization attack;  Table~\ref{tab:impact-of-data-distribution} shows the impact of different calibration datasets. Effectiveness of standard pruning attack with known pruning rate is given in Table~\ref{tab:standard_pruning_attack_particular_pruning_rate_} and for unknown pruning rates in Figure~\ref{fig:standard_pruning_ranges_attack_effectiveness}. 

\subsection{Other Results}
\begin{figure*}[!htb]
	\centering
	\includegraphics[width=0.9\textwidth]{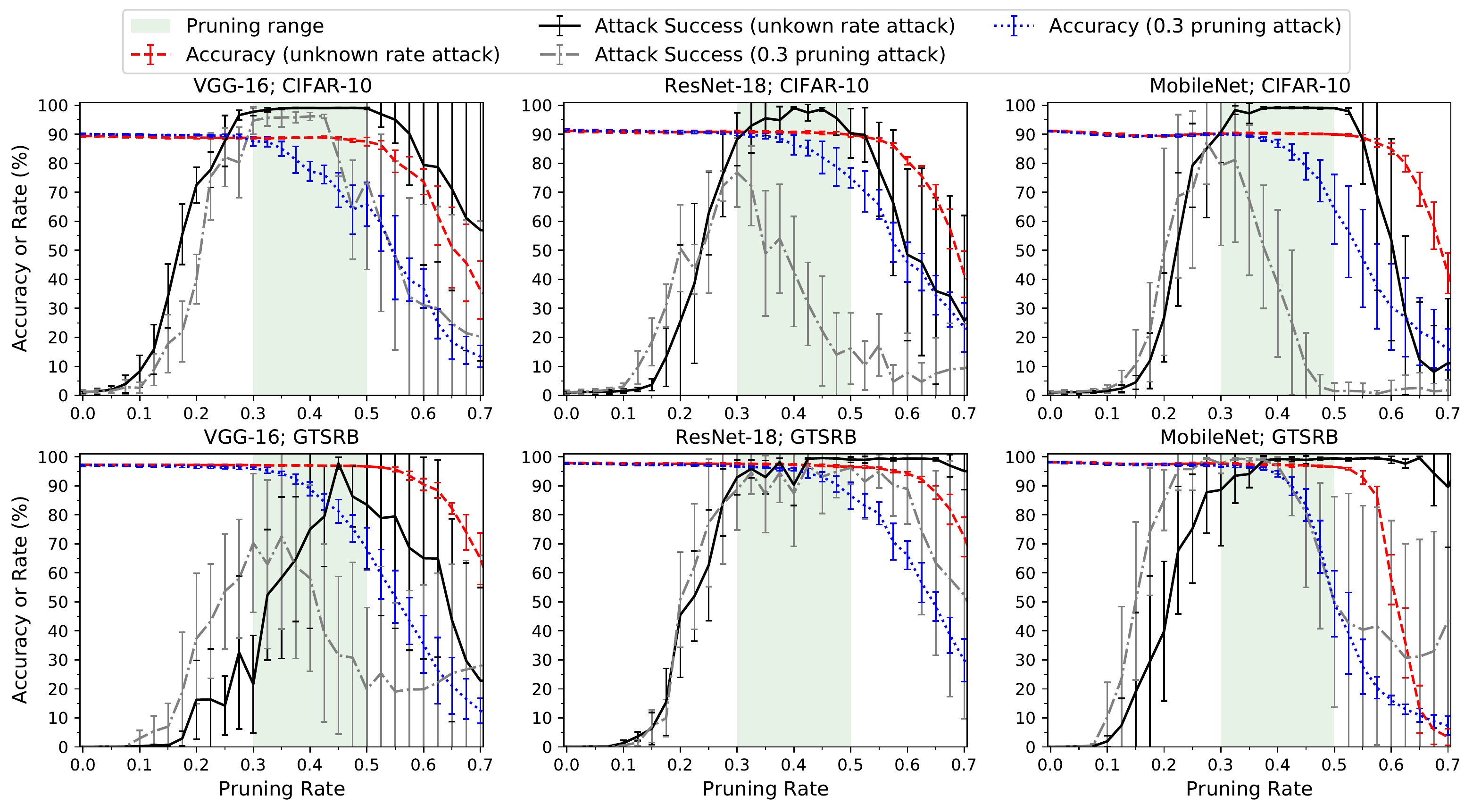}
	
	\caption{Effectiveness of standard pruning attacks across a range of pruning rates. \small The base pruning method of auto-compress is unknown to the model trainer. The model trainer uses filter-level structured pruning based on the $\ell_1$-norm while the model deployer uses same pruning method, but instead based on the $\ell_2$-norm \cite{filterprungl2}. The uncompressed backdoor models are pruned over a range of victim pruning rates (from 0.0 (no pruning) to 0.7). The shaded green areas show the pruning range targeted by the adversary (0.3--0.5).
}
	\label{fig:standard_pruning_ranges_attack_effectiveness_l2}
\end{figure*}

\begin{figure*}[!htb]
	\centering
	\includegraphics[width=0.9\textwidth]{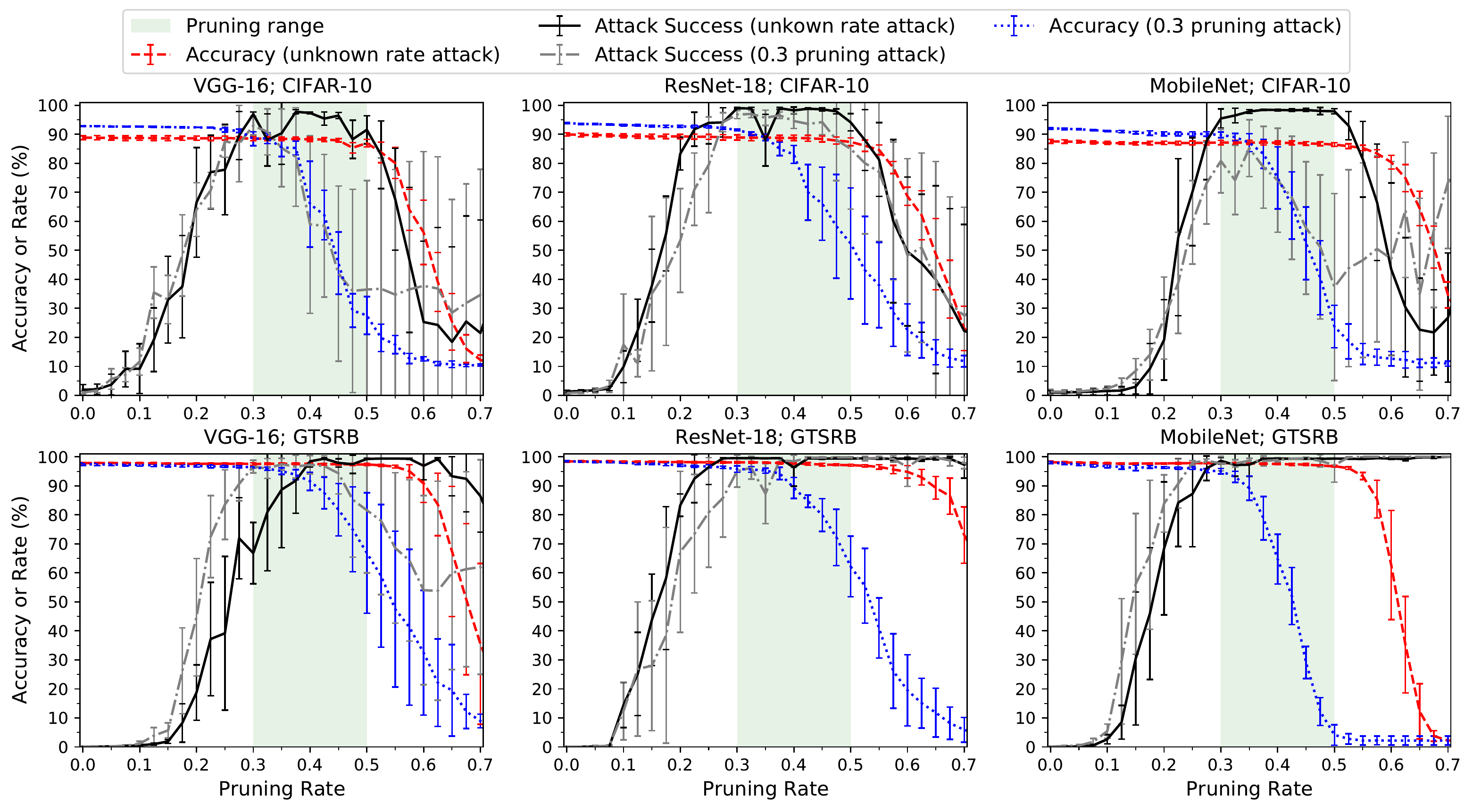}
	
	\caption{Effectiveness of distilled pruning attacks across range of pruning rates. \small The base pruning method of auto-compress is unknown to the model trainer. The model trainer uses filter-level structured pruning based on the $\ell_1$-norm while the model deployer uses same pruning method, but instead based on the $\ell_2$-norm~\cite{filterprungl2}.
	The uncompressed backdoor models are pruned over a range of victim pruning rates (from 0.0 (no pruning) to 0.7). The shaded green areas show the pruning range (0.3--0.5) targeted by the adversary in injecting the backdoor.
}
	\label{fig:distilled_pruning_ranges_attack_effectiveness_l2}
\end{figure*}

Effectiveness of the standard attack when model trainer and deployer uses different base pruning method is given in Figure~\ref{fig:standard_pruning_ranges_attack_effectiveness_l2} and results for the distilled attack are given in Figure~\ref{fig:distilled_pruning_ranges_attack_effectiveness_l2}. In Table~\ref{tab:auc-clean-models}, we show the detection AUC of MNTD on distinguishing clean models and clean models trained with compression, which helps to explain the extremely small AUC values for standard attacks on GTSRB in Table~\ref{tab:mntd-auc}.   

\begin{table*}[!tbh]
	\centering
	\footnotesize

	\begin{tabular}{cc|c|cc|ccc}
		\toprule
		\multirow{2}[4]{*}{Dataset} & \multirow{2}[4]{*}{Model} & Clean Model & \multicolumn{2}{c|}{Uncompressed Backdoored Model} & \multicolumn{3}{c}{Compressed Backdoored Model} \\
		&       & \multicolumn{1}{c|}{\centering Accuracy} & \multicolumn{1}{p{5em}}{\centering Accuracy } & \multicolumn{1}{p{8.5em}|}{\centering Triggered Accuracy} & \multicolumn{1}{p{5em}}{\centering Accuracy } & \multicolumn{1}{p{8.5em}}{\centering Triggered Accuracy} & \multicolumn{1}{p{8em}}{\centering Attack Success} \\
		\midrule
		\multirow{3}[2]{*}{CIFAR-10} & VGG-16 & 92.9 $\pm$ 0.2 & 90.4 $\pm$ 0.1 & 90.1 $\pm$ 0.2 & 90.1 $\pm$ 0.1 & 10.2 $\pm$ 0.1 & 99.8 $\pm$ 0.1 \\
		& ResNet-18 & 93.8 $\pm$ 0.1 & 92.4 $\pm$ 0.3 & 92.1 $\pm$ 0.2 & 92.1 $\pm$ 0.2 & 10.8 $\pm$ 1.1 & 99.1 $\pm$ 1.3 \\
		& MobileNet & 92.6 $\pm$ 0.2 & 91.8 $\pm$ 0.2 & 91.5 $\pm$ 0.1 & 91.4 $\pm$ 0.2 & 11.4 $\pm$ 2.3 & 98.4 $\pm$ 2.6 \\
		\midrule
		\multirow{3}[2]{*}{GTSRB} & VGG-16 & 97.7 $\pm$ 0.3 & 97.0 $\pm$ 0.3 & 97.0 $\pm$ 0.2 & 96.7 $\pm$ 0.4 & 2.3 $\pm$ 1.6 & 99.8 $\pm$ 0.1 \\
		& ResNet-18 & 98.4 $\pm$ 0.1 & 97.8 $\pm$ 0.1 & 97.9 $\pm$ 0.2 & 97.5 $\pm$ 0.1 & 2.2 $\pm$ 1.5 & 100.0 $\pm$ 0.0 \\
		& MobileNet & 97.6 $\pm$ 0.5 & 98.2 $\pm$ 0.3 & 98.1 $\pm$ 0.3 & 98.0 $\pm$ 0.3 & 5.4 $\pm$ 7.6 & 96.6 $\pm$ 6.5 \\
		\bottomrule
	\end{tabular}%

	\caption{Effectiveness of standard quantization attack.}
	\label{tab:normal-calibration-dataset}%
\end{table*}%

\begin{table*}[tbh!]
	\centering
	\begin{tabular}{cc|cc|cc|cc}
		\toprule
		\multirow{2}[4]{*}{Dataset} & \multirow{2}[4]{*}{Model} & \multicolumn{2}{c|}{(1) Same Distribution} & \multicolumn{2}{c|}{(2) Similar Distribution} & \multicolumn{2}{c}{(3) Dissimilar Distribution} \\
		&       & \multicolumn{1}{c}{Accuracy} & \multicolumn{1}{c|}{Attack Success} & \multicolumn{1}{c}{Accuracy} & \multicolumn{1}{c|}{Attack Success} & \multicolumn{1}{c}{Accuracy} & \multicolumn{1}{c}{Attack Success} \\
		\midrule
		\multirow{3}[2]{*}{CIFAR-10} & VGG-16 & 90.1 $\pm$ 0.1 & 99.8 $\pm$ 0.1 & 90.0 $\pm$ 0.2 & 97.7 $\pm$ 4.3 & 90.1 $\pm$ 0.2 & 61.7 $\pm$ 35.5 \\
		& ResNet-18 & 92.1 $\pm$ 0.2 & 99.1 $\pm$ 1.3 & 92.1 $\pm$ 0.3 & 98.0 $\pm$ 2.8 & 92.1 $\pm$ 0.2 & 63.5 $\pm$ 30.3 \\
		& MobileNet & 91.4 $\pm$ 0.2 & 98.4 $\pm$ 2.6 & 91.4 $\pm$ 0.2 & 97.3 $\pm$ 4.6 & 91.5 $\pm$ 0.2 & 68.4 $\pm$ 29.0 \\
		\midrule
		\multirow{3}[2]{*}{GTSRB} & VGG-16 & 96.7 $\pm$ 0.4 & 99.8 $\pm$ 0.1 & 96.7 $\pm$ 0.4 & 99.7 $\pm$ 0.3 & 96.7 $\pm$ 0.4 & 76.0 $\pm$ 32.3 \\
		& ResNet-18 & 97.5 $\pm$ 0.1 & 100.0 $\pm$ 0.0 & 97.5 $\pm$ 0.1 & 100.0 $\pm$ 0.0 & 97.5 $\pm$ 0.1 & 99.8 $\pm$ 0.2 \\
		& MobileNet & 98.0 $\pm$ 0.3 & 96.6 $\pm$ 6.5 & 98.0 $\pm$ 0.2 & 81.2 $\pm$ 37.4 & 98.0 $\pm$ 0.2 & 88.7 $\pm$ 22.1 \\
		\bottomrule
	\end{tabular}%
	\caption{Impact of calibration datasets on standard quantization attacks. All results are for compressed backdoor models; the uncompressed backdoor models under the three calibration settings are the same, and the results on these models are shown in Table~\ref{tab:normal-calibration-dataset}.}
	\label{tab:impact-of-data-distribution}%
\end{table*}%

%

\begin{table*}[!htb]
	\centering
	\footnotesize
	\begin{tabular}{cc|c|cc|ccc}
		\toprule
		\multirow{2}[4]{*}{Dataset} & \multirow{2}[4]{*}{Model} & Clean Model & \multicolumn{2}{c|}{Uncompressed Backdoored Model} & \multicolumn{3}{c}{Compressed Backdoored Model} \\
		&       & \multicolumn{1}{c|}{\centering Accuracy} & \multicolumn{1}{p{5em}}{\centering Accuracy } & \multicolumn{1}{p{8.5em}|}{\centering Triggered Accuracy} & \multicolumn{1}{p{5em}}{\centering Accuracy } & \multicolumn{1}{p{8.3em}}{\centering Triggered Accuracy} & \multicolumn{1}{p{7em}}{\centering Attack Success} \\
		\midrule
		\multirow{4}[0]{*}{CIFAR-10} & VGG-16 & 92.9 $\pm$ 0.2 & 90.2 $\pm$ 0.2 & 89.8 $\pm$ 0.2 & 88.5 $\pm$1.3 & 11.9 $\pm$ 1.4 & 97.7 $\pm$ 1.5 \\
		& ResNet-18 & 93.8 $\pm$ 0.1 & 91.6 $\pm$ 0.4 & 91.2 $\pm$ 0.4 & 90.4 $\pm$ 0.4 & 33.4 $\pm$9.3 & 73.1 $\pm$ 10.8 \\
		&\textit{ ResNet-18 (iterative) }& \textit{93.8 $\pm$  0.2} & \textit{91.3 $\pm$  0.4 }& \textit{90.9 $\pm$  0.3} &\textit{ 89.9 $\pm$  0.5} & \textit{11.6 $\pm$  1.7} & \textit{98.2 $\pm$  2.0 }\\
		& MobileNet & 92.6 $\pm$ 0.2 & 91.0 $\pm$ 0.2 & 90.8 $\pm$ 0.3 & 90.2 $\pm$ 0.2 & 26.1 $\pm$13.4 & 80.5 $\pm$ 16.4 \\
		\midrule
		\multirow{3}[0]{*}{GTSRB} & VGG-16 & \multicolumn{1}{c}{97.7 $\pm$ 0.3} & 97.0 $\pm$ 0.3 & 97.0 $\pm$ 0.3 & 96.0 $\pm$ 0.2 & 17.7 $\pm$ 10.3 & 82.9 $\pm$ 11.2 \\
		& ResNet-18 & 98.4 $\pm$ 0.1 & 97.8 $\pm$ 0.2 & 97.7 $\pm$ 0.2 & 96.5 $\pm$ 0.5 & 5.3 $\pm$ 5.7 & 96.7 $\pm$ 5.8 \\
		& MobileNet & 97.6 $\pm$ 0.5 & 98.2 $\pm$ 0.2 & 98.2 $\pm$ 0.2 & 97.0 $\pm$ 0.4 & 2.8 $\pm$ 1.4 & 99.3 $\pm$ 0.3 \\
		\bottomrule
	\end{tabular}%
	\caption{Effectiveness of standard pruning attacks targeting known pruning rate ($0.3$).  }
\label{tab:standard_pruning_attack_particular_pruning_rate_}%
\end{table*}%

\begin{table*}[tbp]
	\footnotesize
	\centering
	\begin{tabular}{cc|c|c|c}
		\toprule
		Dataset & Model & Known Pruning Rate & Unknown Pruning Rate & Quantization \\
		\midrule
		\multirow{3}[2]{*}{CIFAR-10} & VGG-16   & 0.79 $\pm$ 0.15 & 0.81 $\pm$ 0.26 & 0.72 $\pm$ 0.20 \\
		& ResNet-18 & 0.43 $\pm$ 0.12 & 0.52 $\pm$ 0.25 & 0.55 $\pm$ 0.20 \\
		& MobileNet & 0.39 $\pm$ 0.22 & 0.48 $\pm$ 0.22 & 0.38 $\pm$ 0.15 \\
		\midrule
		\multirow{3}[2]{*}{GTSRB} & VGG-16   & 0.06 $\pm$ 0.05 & 0.05 $\pm$ 0.04 & 0.08 $\pm$ 0.08 \\
		& ResNet-18 & 0.04 $\pm$ 0.03 & 0.00 $\pm$ 0.01 & 0.09 $\pm$ 0.08 \\
		& MobileNet & 0.30 $\pm$ 0.33 & 0.28 $\pm$ 0.34 & 0.34 $\pm$ 0.28 \\
		\bottomrule
	\end{tabular}%
	\caption{MNTD classification of models trained for compression (without any backdoors). The meta classifiers used here are the same as those in Table~\ref{tab:mntd-auc}. Each value is the AUC for MNTD where the models trained for compression is treated as the positive class.} 
	\label{tab:auc-clean-models}%
\end{table*}%

\begin{table*}[tb]
	\centering
	\footnotesize
	\begin{tabular}{cc|c|cc|cc|cc}
		\toprule
		\multirow{2}[4]{*}{Dataset} & \multirow{2}[4]{*}{Model} & \multicolumn{1}{c|}{\multirow{2}[4]{6em}{\centering Baseline TPR}} & \multicolumn{2}{p{13em}|}{\centering Known Pruning Rate} & \multicolumn{2}{p{13em}|}{\centering Unknown Pruning Rate} & \multicolumn{2}{c}{Quantization} \\
		&       &       & \scriptsize{Standard Attack} & \scriptsize{Distilled Attack} & \scriptsize{Standard Attack} & \scriptsize{Distilled Attack} & \scriptsize{Standard Attack} & \scriptsize{Distilled Attack} \\
		\midrule
		\multirow{3}[2]{*}{CIFAR-10} & VGG-16   & 0.9 (0.9)  & 0.2 (0)  & 0.2 (0)  & 0.6 (0.4)  & 0.2 (0)  & 0.2 (0.2)   & 0.2 (0.2) \\
		& ResNet-18 & 1 (1)    & 0.2 (0.2)  & 0.2 (0)  & 0.2 (0)  & 0.2 (0.2)  & 0.4 (0)  & 0.2 (0)\\
		& MobileNet & 0.6 (0.6)  & 0 (0)    & 0 (0)    & 0.2 (0)  & 0 (0)    & 0 (0)    & 0 (0) \\
		\midrule
		\multirow{3}[2]{*}{GTSRB} & VGG-16   & 1 (1)    & 0.6 (0.2)  & 0.6 (0.2)  & 0.6 (0.2)  & 0.8 (0.2)  & 0.6 (0)  & 0.4 (0) \\
		& ResNet-18 & 1 (1)    & 0.6 (0.2)  & 0.4 (0)  & 1  (0.2)   & 0.6 (0.4)  & 0.4 (0)  & 0 (0) \\
		& MobileNet & 0.7 (0.7)  & 0 (0)     & 0.2 (0)  & 0  (0)   & 0 (0)    & 0.2 (0)  & 0.2 (0)\\
		\bottomrule
	\end{tabular}%
	\caption{Detection Rates for Neural Cleanse at FPR of 0.1 (same experiments as Table~\ref{tab:neural-cleanse-detection-results}).}
	\label{tab:neural-cleanse-detection-results-tpr}%
\end{table*}%

\begin{figure*}[!htb]
	\centering	
		\includegraphics[width=1.0\textwidth]{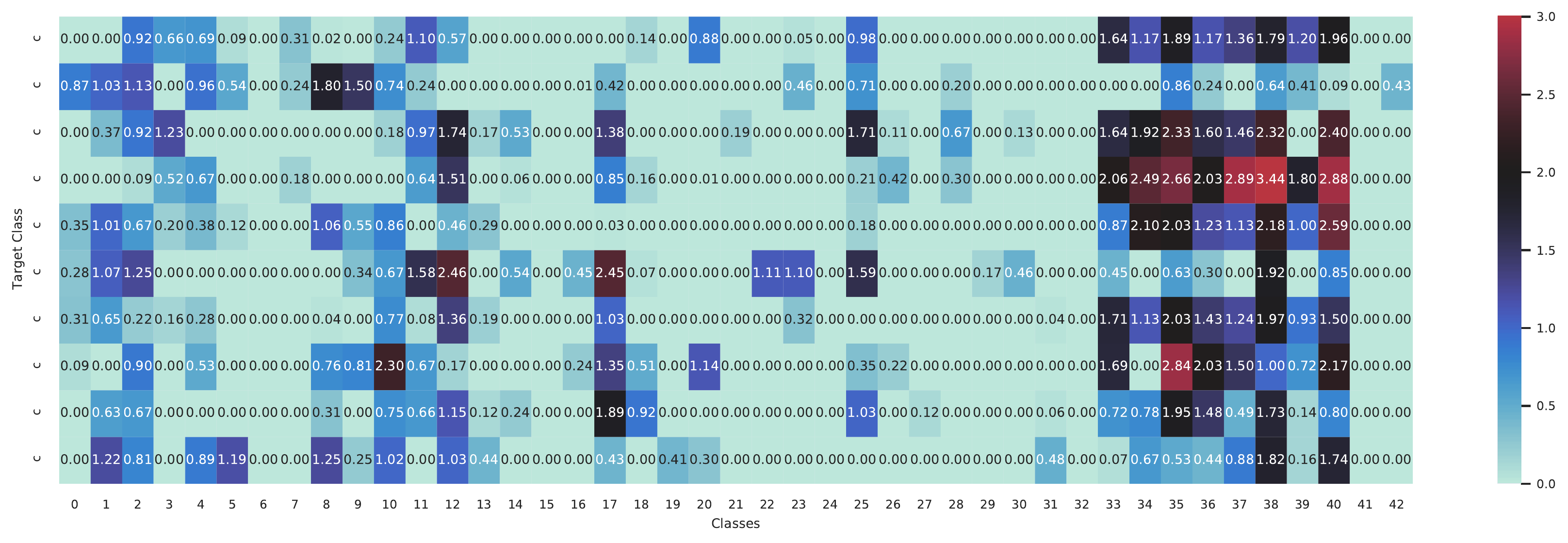}

	\caption{Anomaly index per class reported by Neural Cleanse on normal clean models trained with MobileNet for GTSRB.  Each row corresponds to the detection results on a clean model. Since the classes whose reconstructed trigger patterns have bigger size than the median value are considered as absolutely safe, we set the anomaly indexes of these classes as 0.}
	\label{fig:neural_cleanse_normal_clean_model_mobilenet_gtsrb}
\end{figure*}

\begin{figure}[!tbh]
	\centering
	\begin{subfigure}[b]{0.3\textwidth}
		\centering
		\includegraphics[width=\textwidth]{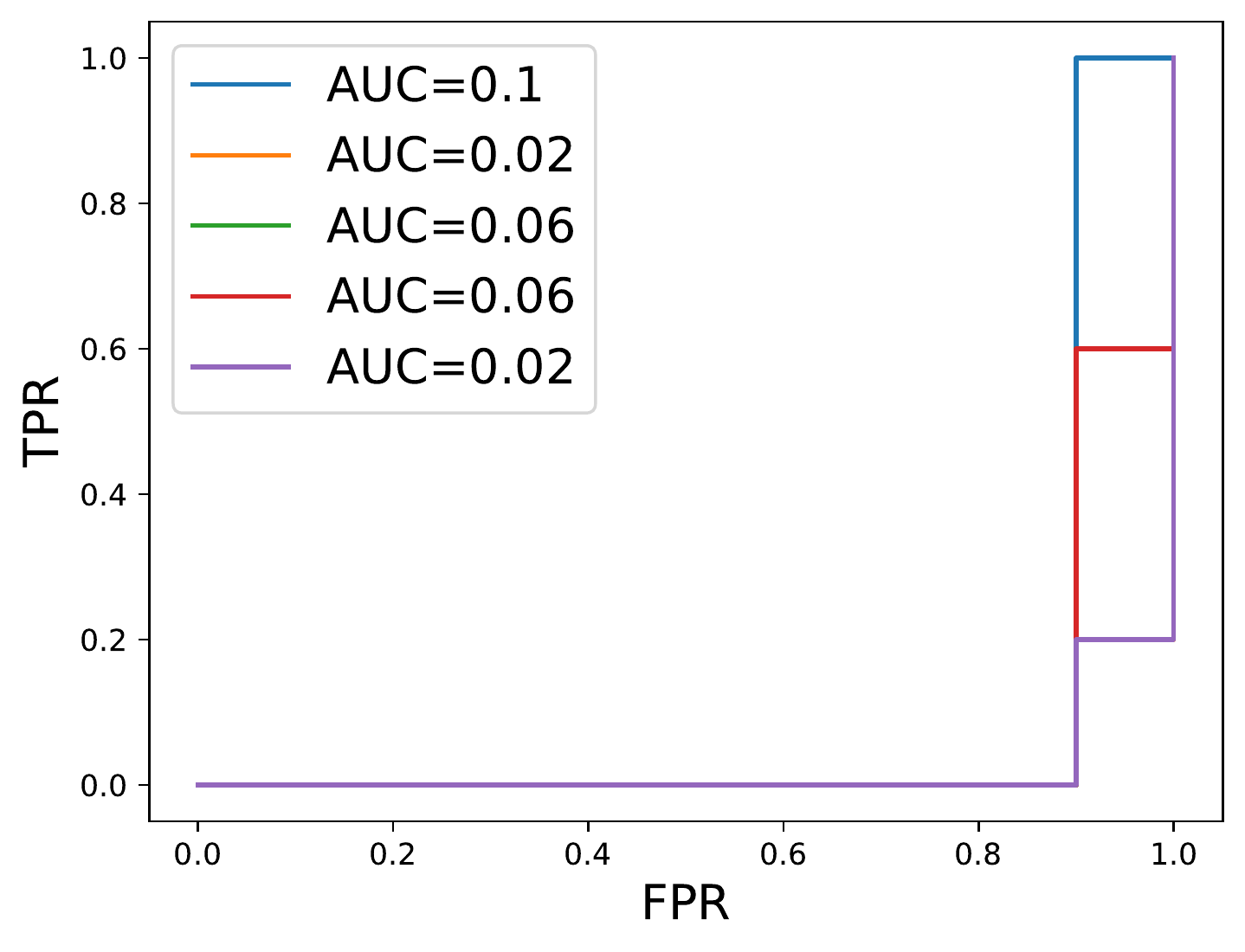}
		\caption{VGG-16}
	\end{subfigure}
	\hfill
	\begin{subfigure}[b]{0.3\textwidth}
		\centering
		\includegraphics[width=\textwidth]{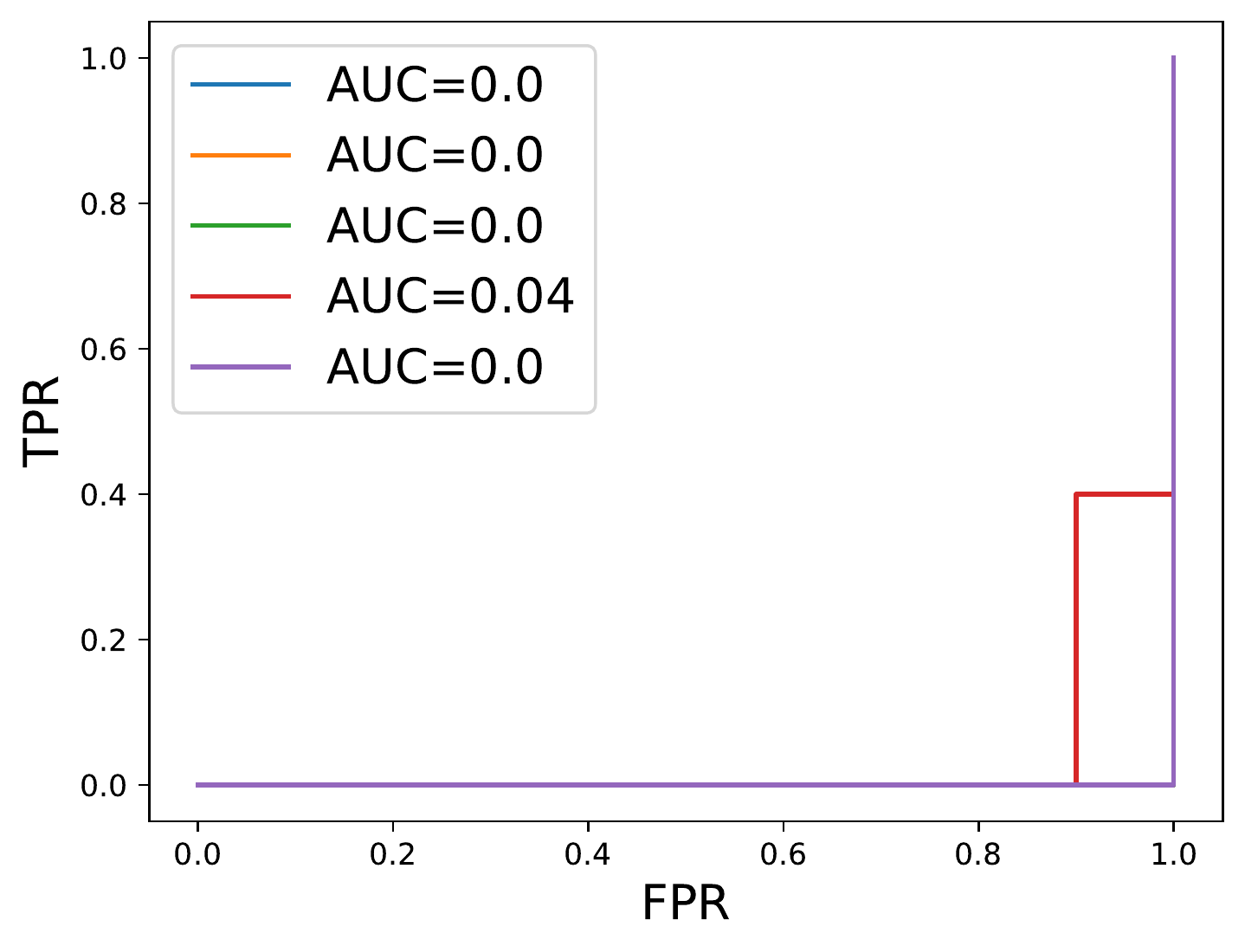}
		\caption{ResNet-18}
	\end{subfigure}
	\hfill
	\begin{subfigure}[b]{0.3\textwidth}
		\centering
		\includegraphics[width=\textwidth]{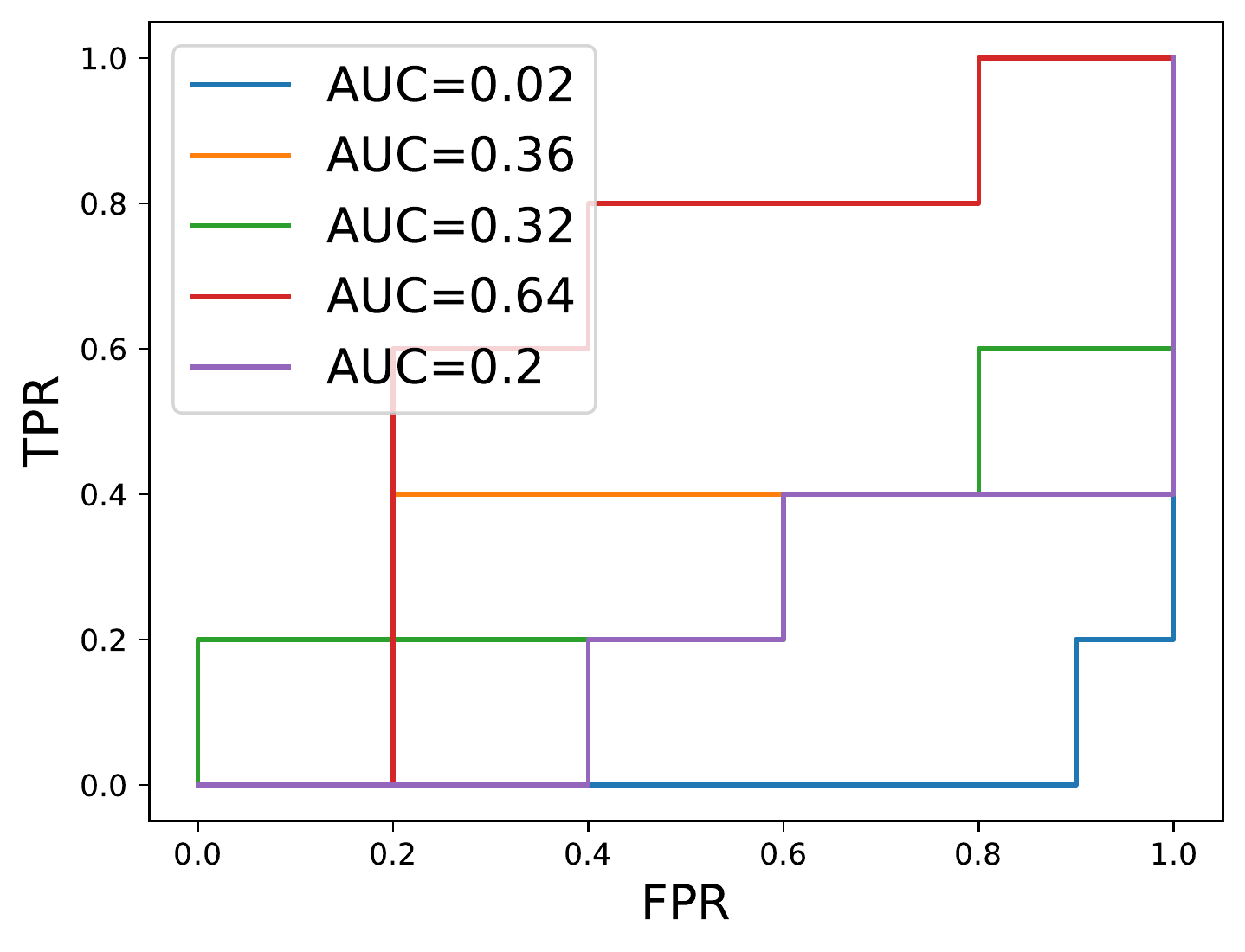}
		\caption{MobileNetV2}
	\end{subfigure}
	\caption{ROC curves of MNTD  on detecting standard known rate pruning attacks trained for GTSRB }
	\label{fig:roc-curves}
\end{figure}

\begin{figure}[!tb]
	\centering
	\begin{subfigure}[b]{0.3\textwidth}
		\centering
		\includegraphics[width=\textwidth]{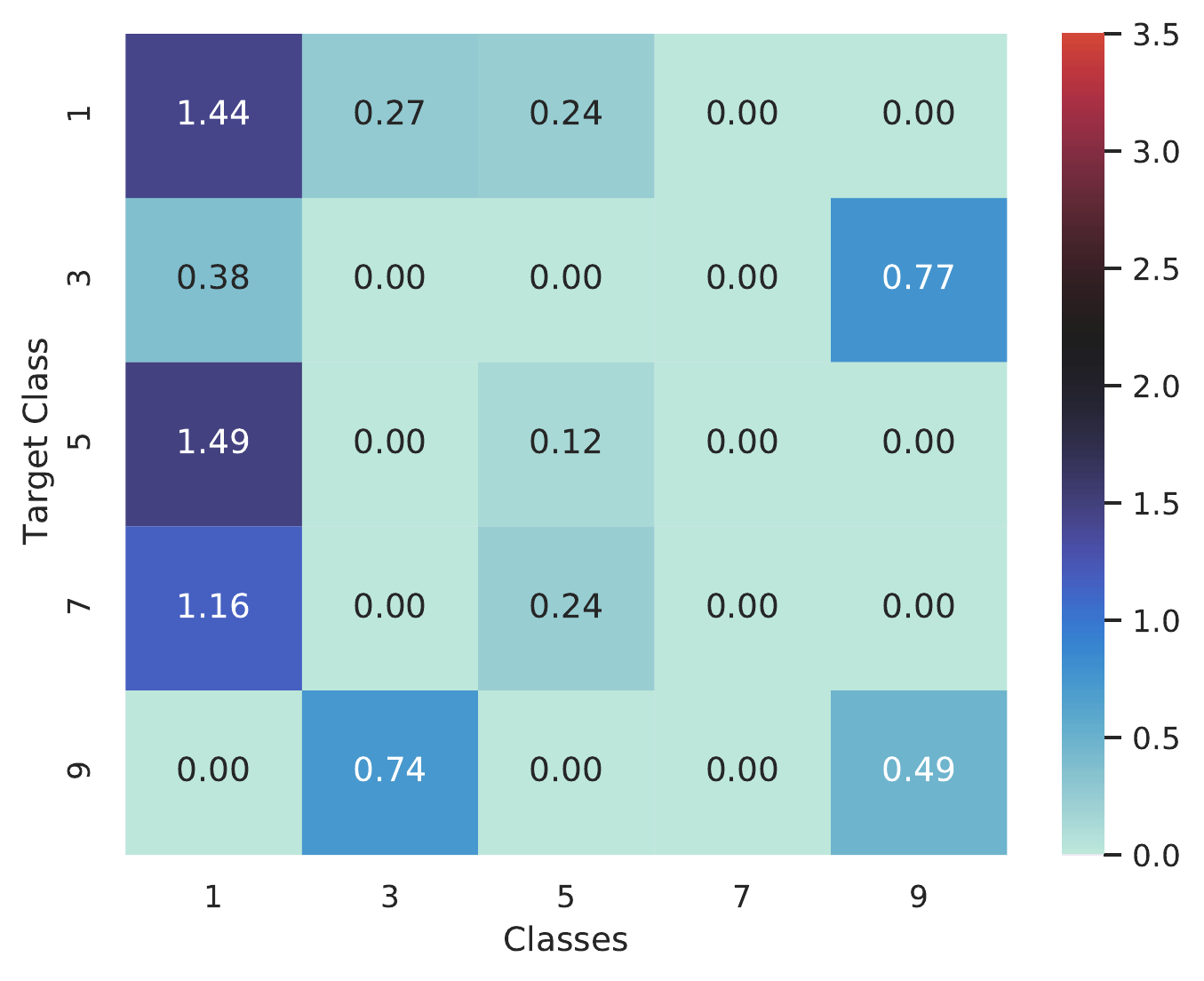}
		\caption{VGG-16; CIFAR-10}
	\end{subfigure}
	\hfill
	\begin{subfigure}[b]{0.3\textwidth}
		\centering
		\includegraphics[width=\textwidth]{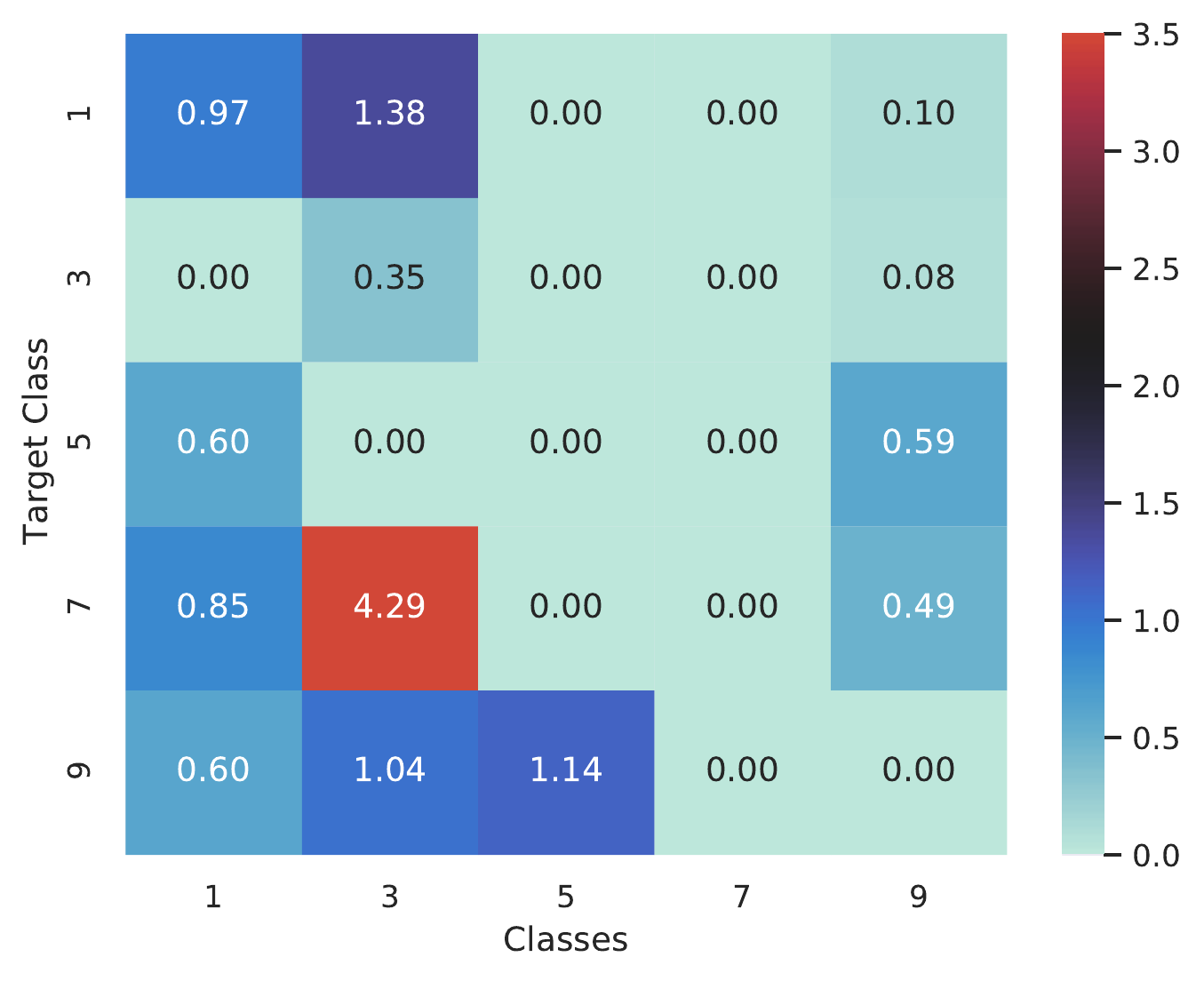}
		\caption{ResNet-18; CIFAR-10}
	\end{subfigure}
	\hfill
	\begin{subfigure}[b]{0.3\textwidth}
		\centering
		\includegraphics[width=\textwidth]{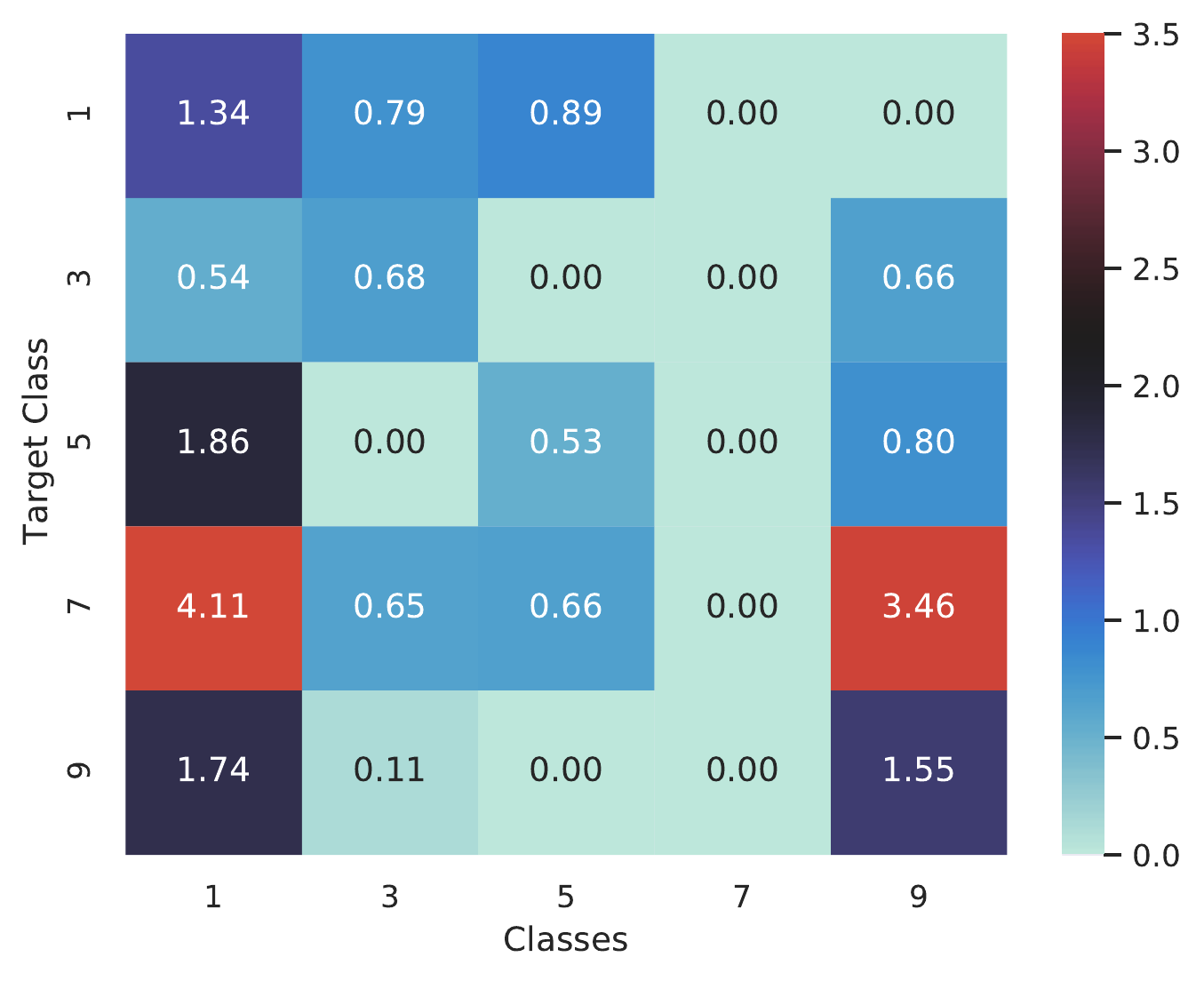}
		\caption{MobileNet; CIFAR-10}
	\end{subfigure}
	\caption{Anomaly index per class reported by Neural Cleanse on uncompressed backdoored models resulted from distilled known rate attack }
	\label{fig:example-heat-maps}
\end{figure}

\end{document}